\documentclass[10pt,twocolumn, aps, pra, superscriptaddress,floatfix]{revtex4}
\usepackage{graphicx}
\usepackage[english]{babel}
\usepackage{amsmath}
\usepackage{amssymb}
\usepackage[usenames,dvipsnames]{color}
\usepackage{color}
\definecolor{Blue}{rgb}{0.3,0.3,0.9}
\definecolor{orange}{rgb}{1,0.5,0}
\usepackage{subfigure}
\usepackage{bm}
\usepackage{titlesec}
\usepackage{mathrsfs}
\usepackage{esint}
\usepackage[unicode=true,  bookmarks=false,  breaklinks=false,pdfborder={0 0  1},backref=false,colorlinks=true]  {hyperref}

\newcommand{\sgn}{{\mbox{sgn}}}
\newcommand{\im}{{\mbox{Im}}}

\newcommand{\amp}{{\mbox{Amp}}}

\newcommand{\bx}{{\mathbf x}}
\newcommand{\by}{{\mathbf y}}
\newcommand{\bz}{{\mathbf z}}
\newcommand{\bw}{{\mathbf w}}
\newcommand{\bp}{{\mathbf p}}
\newcommand{\bq}{{\mathbf q}}
\newcommand{\bk}{{\mathbf k}}

\newcommand{\sx}{S^x}

\newcommand{\sz}{S^z}

\newcommand{\s}{\sigma}

\begin{document}

\author{Mohammad F. Maghrebi}
\email{magrebi@umd.edu}

\affiliation{Joint Quantum Institute, NIST/University of Maryland, College Park, Maryland 20742, USA}

\affiliation{Joint Center for Quantum Information and Computer Science, NIST/University of Maryland, College Park, Maryland 20742, USA}

\author{Zhe-Xuan Gong}

\affiliation{Joint Quantum Institute, NIST/University of Maryland, College Park, Maryland 20742, USA}

\affiliation{Joint Center for Quantum Information and Computer Science, NIST/University of Maryland, College Park, Maryland 20742, USA}

\author{Michael Foss-Feig}

\affiliation{Joint Quantum Institute, NIST/University of Maryland, College Park, Maryland 20742, USA}

\affiliation{Joint Center for Quantum Information and Computer Science, NIST/University of Maryland, College Park, Maryland 20742, USA}

\author{Alexey V. Gorshkov}

\affiliation{Joint Quantum Institute, NIST/University of Maryland, College Park, Maryland 20742, USA}

\affiliation{Joint Center for Quantum Information and Computer Science, NIST/University of Maryland, College Park, Maryland 20742, USA}

\title{Causality and quantum criticality in long-range lattice models}
\begin{abstract}
  Long-range quantum lattice systems often exhibit drastically different behavior than their short-range counterparts. In particular, because they do not satisfy the conditions for the Lieb-Robinson theorem, they need not have an emergent relativistic structure in the form of a light cone.  Adopting a field-theoretic approach, we study the one-dimensional transverse-field Ising model with long-range interactions, and a fermionic model with long-range hopping and pairing terms, explore their critical and near-critical behavior, and characterize their response to local perturbations. We deduce the dynamic critical exponent, up to the two-loop order within the renormalization group theory, which we then use to characterize the emergent causal behavior. We show that beyond a critical value of the power-law exponent of the long-range couplings, the dynamics effectively becomes relativistic. Various other critical exponents describing correlations in the ground state, as well as deviations from a linear causal cone, are deduced for a wide range of the power-law {exponent}.

\end{abstract}

\maketitle

\begin{table*}[!htb]
    \begin{minipage}{.5\linewidth}
      \centering
{\bf TFIM}\\
\null
\begin{tabular}{|c||c|c|c|} \hline
&
\quad $0<\s<\frac{2}{3}$ \quad \null  & $\frac{2}{3} < \s < \frac{7}{4}$ & \quad $\frac{7}{4}<\s \quad \null $ \\ \hline\hline
\quad $z$  \qquad &
\begin{tabular}{c}
\\
$\s/2$
\\
\\
\end{tabular} & \quad $\s/2+ {\varsigma}(\s)  \epsilon^2 +{\cal O}(\epsilon^3)$
\quad &
\begin{tabular}{c}
\\
1
\\
\\
\end{tabular}  \\ \hline
\quad $\theta$  \qquad &
\begin{tabular}{c}
\\
$1-\s/2$
\\
\\
\end{tabular} & \quad $1-\s/2+ {\varsigma}(\s)  \epsilon^2 +{\cal O}(\epsilon^3)$
\quad &
\begin{tabular}{c}
\\
1/4
\\
\\
\end{tabular}  \\ \hline
\end{tabular}
    \end{minipage}%
    \begin{minipage}{.5\linewidth}
      \centering
        {\bf Fermion Model}\\
\null
\begin{tabular}{|c||c|c|} \hline
& \quad $0<\s<1$ \quad \null  & \quad $1 < \s $   \quad \\ \hline\hline
\quad $z$  \qquad &
\begin{tabular}{c}
\\
$\s$
\\
\\
\end{tabular} & \begin{tabular}{c}
 \\
    \quad 1\quad \null
      \\
   \\
\end{tabular}
\\ \hline
\quad $\theta$  \qquad &
\begin{tabular}{c}
{$2-{\s}$} \\
\end{tabular}
&
\begin{tabular}{c}
 \\
    \quad 1\quad \null
      \\
   \\
\end{tabular} \\ \hline
\end{tabular}
\end{minipage}
\caption{The exponents describing two long-range lattice models at criticality: TFIM and Fermion Model.
The long-range interaction between Ising spins or long-range hopping and pairing terms in the fermionic model is assumed to fall off as $\sim 1/R^{1+\sigma}$
where $\s>0$. The exponent $z$ denotes the dynamic critical exponent; $z=1$ defines a linear light cone, while $z<1$ corresponds to a sublinear causal region. The exponent $\theta$ characterizes the decay of the correlation function in the ground state at criticality ($\theta$ being the anomalous dimension $\eta$ reported in Sec.~\ref{Sec: Ising Model} for the TFIM). For the TFIM, and in an intermediate range of the power-law exponent, the critical exponents should be computed in a series of epsilon-expansion with $\epsilon=3\s/2-1$ and ${\varsigma}(\s)$ a rather complicated expression approximately given by ${\varsigma}(\s)\approx 1/[24(1+\s^2)]$.
Even away from the critical point,
the correlation functions decay as a power law $\sim 1/R^{1+\s}$ (not shown in this table).
}\label{Tab I}
\end{table*}

\begin{table*}[!htb]
    \begin{minipage}{.5\linewidth}
      \centering
      {\bf TFIM}\\
\null
\begin{tabular}{|c||c|c|c|} \hline
& $0<\s<1$ \null  & $1 < \s < \frac{7}{4}$ & \quad $\frac{7}{4}<\s \quad \null $ \\ \hline\hline
 Critical   &
\multicolumn{1}{r}{}& \begin{tabular}{c}
 \hskip -50pt Non-linear
\\
\hskip -50pt
$D_\s(t,R)={R^{-\theta}}\,\,g_\s\!\left({t}/{R^z}\right)$
\\
\end{tabular} &
\begin{tabular}{c}
\\ Linear
\\
\\
\end{tabular} \\ \hline \hline
 Non-critical &
\begin{tabular}{c}
\end{tabular}
Non-linear
&\multicolumn{1}{r}{}
\begin{tabular}{c}
 \\
   \\
   \\
   \\
\end{tabular} &
\begin{tabular}{c}
 \hskip -40pt Linear
  \\
\end{tabular} \\ \hline
\end{tabular}
    \end{minipage}%
    \begin{minipage}{.5\linewidth}
      \centering
            {\bf Fermion Model}\\
        \null
        \begin{tabular}{|c||c|c|c|} \hline
&  $0<\s<\frac{1}{2}$ \null  & $\frac{1}{2} < \s <1$ \null& \quad $1<\s \quad \null $ \\ \hline\hline
Critical  &
\multicolumn{1}{r}{}& \begin{tabular}{c}
 \hskip -50pt Non-linear
\\ \hskip -50pt
${\mathbf D}_{\s}(t,R)={R}^{-1}\,\,{\mathbf H}_\s\!\left({t}/{R^{z}}\right)$
\\
\end{tabular} &
\begin{tabular}{c}
\\
Linear
\\
\\
\end{tabular} \\ \hline \hline
\begin{tabular}{c}
Non-critical \\ (Hopping)
\end{tabular} &
 \multicolumn{1}{r}{}
\begin{tabular}{c}
   \\
   \\
\end{tabular} &
\begin{tabular}{c}
\end{tabular}
\hskip -80pt Non-linear
&
\begin{tabular}{c}
  Linear
  \\
\end{tabular} \\ \hline
\begin{tabular}{c}
Non-critical  \\ (Pairing)
\end{tabular}
&
\begin{tabular}{c}
\end{tabular}
Non-linear
&\multicolumn{1}{r}{}
\begin{tabular}{c}
 \\
   \\
\end{tabular} &
\begin{tabular}{c}
 \hskip -50pt Linear
\end{tabular} \\ \hline
\end{tabular}
    \end{minipage}
       \caption{The causal behavior of long-range TFIM and Fermion Model at or away from criticality. For long-range models with a sufficiently rapidly decaying power-law, the causal behavior is described by a linear light cone.
       For exponents smaller than a critical value, which depends on the specific model, the causal behavior is not linear. This critical value of the power-law exponent, for the models at criticality, is given by the onset of the dynamic critical exponent deviating from $z=1$, cf. Tab.~\ref{Tab I}. At criticality, the response functions $D_\s$ (TFIM) and ${\mathbf D}_{\s}$ (Fermion Model) are described by the critical exponents $z$ and $\theta$, and the general scaling functions $g_\s$ (TFIM) and ${\mathbf H}_\s$ (Fermion Model) as detailed in the text. For the critical TFIM model with  $\sigma>7/4$, and the fermionic model with $\sigma>1$, the response function describes a linear light cone.
       Away from criticality, linear/non-linear behaviors are predicted by a simple analysis of the dispersion relation. Analytical expressions describing various regimes are provided in the text. For the noncritical Fermion Model, long-range hopping or pairing cases exhibit different causal behavior.
       }\label{Tab II}
\end{table*}

\section{Introduction and Summary}
Long-range interactions arise in a wide range of physical systems. Examples include NV centers and other solid-state defects \cite{childress06,balasubramanian09,weber10,dolde13}, excitons (Frenkel excitations) in organic solids \cite{agranovich09}, polarons \cite{Alexandrov95}, Shiba chains \cite{Pientka13,Pientka14}, and photon-mediated interactions between superconducting qubits \cite{Houck12}.
Furthermore, long-range interactions emerge very naturally---and are often unavoidable---in atomic, molecular, and optical (AMO) systems, for example van der Waals ($1/R^6$) interactions between Rydberg atoms \cite{saffman10,schauss12} or polaritons \cite{firstenberg13}, magnetic or electric dipole-dipole ($1/R^3$) interactions between atoms or molecules \cite{saffman10,yan13,aikawa12,lu12,childress06,balasubramanian09,weber10,dolde13}, and variable-range ($1/R^\alpha$) interactions between atoms in multimode
cavities \cite{gopalakrishnan11} or trapped ions \cite{islam13,britton12,Richerme14,Jurcevic14}.
While the quantum-critical behavior of short-range interacting models has been extensively studied, quantum criticality and phase transitions and their universal properties are less fully explored in the presence of long-range interactions.

An important difference between short- and long-range interacting systems, namely the nature of any emergent causal structure, can be characterized by their response to a local perturbation.  The celebrated Lieb-Robinson bound demonstrates that even nonrelativistic quantum systems exhibit a linear `light cone' bounding a causal region, outside of which the \emph{response function} is exponentially suppressed \cite{lieb72}, provided the interactions are short-ranged. This bound enforces the emergence of a `relativistic' causal behavior even in condensed matter systems.  For long-range power-law interactions, on the other hand, the light cone may be sublinear, and the bounds on the influence of a local perturbation are much less stringent. In an early work by Hastings and Koma, the boundary of this region is shown to be at least logarithmic rather than linear \cite{hastings05}.  Recent works have further explored the causal consequences of long-range interactions \cite{hauke13,eisert13,gong14,Rajabpour15,Jurcevic15}, and Ref.~\cite{Foss-Feig15} has recently improved the Hastings-Koma bound by constraining the causal region algebraically rather than logarithmically.
However, tighter bounds are not ruled out, and it remains an open question whether a linear light cone emerges for generic power-law interacting models beyond some critical power-law exponent.

In this paper, we study the causal structure of one-dimensional long-range lattice models in the vicinity of a quantum critical point, and adopt a field-theoretical approach, complementary to the extensive literature on Lieb-Robinson-type bounds for long-range lattice models. Specifically, we compute correlation functions and causal response functions in the ground states of continuum field theories governing the near-critical behavior of the long-range interacting transverse-field Ising model (TFIM), and also a fermionic model with long-range hopping and pairing terms. We identify the so-called dynamic critical exponent that defines the relative scaling of space and time coordinates, which characterizes the causal structure of the underlying lattice model close to its critical point. We show that linear light cones emerge above critical values of the power-law exponent of the long-range couplings, which depend on the spin/fermion model as well as whether the model is at or away from criticality. In both cases, we also identify the critical exponent characterizing the decay of correlations in the ground state. Furthermore, it is shown in detail that the response to a local perturbation obtains a general scaling form, which is determined by the value of these two exponents. Note that approximate numerical approaches to many-body models with long-range interactions \cite{Lode12, Fischer15,Chitra00,Georges96} also exist (in the context of ultracold systems or elsewhere).  In contrast, our field-theory treatment is well suited to extracting universal aspects of the long-distance and long-time behavior of the many-body system, and specifically the critical exponents (beyond their mean-field values) in the thermodynamic limit, which are usually difficult to access via numerical methods.

It is important to note that Lieb-Robinson-type bounds are generally state-independent and agnostic to many details of the underlying model, requiring only a lattice with a finite-dimensional local Hilbert space on each site \cite{lieb72}, and an interaction with some prescribed spatial decay. While these bounds are very general, naive applications to long-range interacting models result in unphysically large causal regions, and obtaining even qualitatively tight rigorous bounds remains an open problem.

In this paper, by limiting ourselves to ground states of specific models and local quenches on such ground states, we find a rich structure not captured in the more general Lieb-Robinson statements. Crucially, our results provide explicit examples of models with particular sub-linear causal regions.  Thus, while existing Lieb-Robinson bounds constrain the speed with which information can propagate from above, our results place a lower limit on how far (towards lower speeds) such upper bounds can ultimately be pushed.  One might expect that the availability of low-energy excitations at a quantum critical point should facilitate faster-than-light propagation of the response to local perturbations in the presence of long-range couplings; we confirm this intuition by comparing critical and noncritical regimes. In this sense, our study of critical points also lends some heuristic support to the conjecture that our results may actually coincide with the best possible Lieb-Robinson bounds for long-range interacting systems.  However, this is certainly speculative; our treatment is complimentary to the Lieb-Robinson type approach and provides some intuition for how the most refined versions of such bounds \emph{might} look, but does not constitute a rigorous result on how such bounds \emph{must} look.

We also stress that our definition of a causal region and linear/non-linear causal behavior is not \emph{necessarily} precisely equivalent to that used in the context of Lieb-Robinson bounds and information theory.  In the latter context, the shape of the causal region is determined by requiring that the Lieb-Robinson bound falls below some threshold value outside of it, which naturally defines a region in space and time within which appreciably large signals can, in principle, be sent.  In our definition, the light-cone shape is tied to the dynamic critical exponent, which is a natural and standard identification from the point of view of field theory. We show that this exponent controls how fast information reaches a given point by identifying it with the space-time scaling of the first local maximum of the response function, and thus there is clearly a close connection to the information-theoretic definition.  We find this definition necessary to extract a light-cone shape from a specific model, since in the absence of perfectly ballistic transport the height of this maximum generically decays in space and time, and constant contours do not extend asymptotically to large distances and times.  This issue presents itself for short-range interacting models as well, where in general the proper light cone shape is only obtained by ignoring the decay of the peak (local maximum) \cite{bose07}, with the understanding that the absence of perfectly ballistic dynamics is a model-specific phenomenon and should not influence the Lieb-Robinson bound.  However, whether this notion exactly coincides with that of information theory is an interesting question in need of further investigation.

We briefly discuss the methods and quantities of interest relevant to our investigation of critical and dynamical aspects of long-range interacting lattice models in Sec.~\ref{Sec. Methods and quantities}. We present our results for the long-range interacting TFIM in Sec.~\ref{Sec: Ising Model}, and for the fermionic model with long-range hopping and pairing terms in Sec.~\ref{Sec: Fermions}. We have studied in detail (i) the correlation/response functions, (ii) for spin/fermion models, (iii) at/away from criticality, and (iv) for different ranges of the exponent characterizing the power-law couplings. In all these models, we are always sufficiently close to criticality that the correlation length is large compared to the lattice spacing, and a continuum description is valid.
For the benefit of the reader, we have summarized our main results in Tables~\ref{Tab I} and \ref{Tab II}. We have performed a detailed renormalization-group (RG) calculation up to the two-loop order for the
TFIM, and argued, on the basis of RG, that the critical exponents obtained in the fermionic model are exact (i.e., mean-field exponents are exact, and would not receive corrections from interactions). For both models, short-range interactions give rise to a dynamical exponent $z=1$, which indicates a linear light cone and relativistic dynamics. We explore in detail how sufficiently slow-decaying power-law couplings can give rise to sublinear light cones, with $z<1$.

\section{Methods and quantities of interest}\label{Sec. Methods and quantities}
In this work, we rely heavily on scaling and renormalization group theory, which provides a systematic way to integrate out short-wavelength degrees of freedom (at the scale of lattice spacing, for example) in order to find an effective description of the physics at long wavelengths. We often find that general quantities of interest take a simple scaling form involving universal exponents that determine how fast correlations fall with distance, or how space and time coordinates scale with respect to each other, see Tables \ref{Tab I} and \ref{Tab II}.Such exponents can be approximately determined (at the mean field level) via a simple power counting, which is the first step of a systematic RG treatment. The knowledge of mean-field exponents can be then used to determine whether interactions affect the universal behavior of the system. In fact, it can be argued, as we will often do, that many types of interaction become less and less important at long wavelengths (the corresponding coefficients are suppressed along the RG flow). In this case, they do not affect the universal behavior, and we say that interactions are not relevant in the sense of RG. In other cases where interactions are relevant, we systematically use renormalization group theory to determine the critical exponents beyond their mean-field values. This approach is particularly appealing since even a complicated theory can be efficiently described by a small set of exponents.

In the remainder of this section we define the various quantities of interest reported in Tables \ref{Tab I} and \ref{Tab II}, separating them into two subsections based on relevance to the TFIM or the fermionic model.  We will focus on properties of these quantities specifically in the limit of long times and large distances.

{\it TFIM.---}In the model studied in Sec.~\ref{Sec: Ising Model}, we shall focus on the $z$-component of the spin, $S^z_i$, that defines the Ising order parameter ($i$ denoting the lattice site in a 1D chain). We are primarily interested in universal properties of correlation functions and causal response functions,
\begin{align}
 iG_\s(t, i-j)&=\langle {\cal T}S^z_i(t)S^z_j(0)\rangle, \label{Eq. G 11}\\
 iD_\s(t,i-j)&=\Theta(t) \langle [S^z_i(t), S^z_j(0)]\rangle \label{Eq. D 11},
\end{align}
where the operator $S^z_i(t)$ is defined in the Heisenberg picture. The subscript $\s$ denotes the exponent of the long-range interaction potential $V(i-j)\sim 1/|i-j|^{1+\s}$; with nearest-neighbor terms only, we have $\s=\infty$.  In the above equations, $\Theta$ is the Heaviside step function, ${\cal T}$ is the time-ordering operator, and the expectation values are computed in the ground state of the system Hamiltonian (see Sec.~\ref{Sec: Ising Model} for specific details). We have also used the translation symmetry to write the two-point functions as a function of the distance (in units of lattice constant) between the two points and their time difference. The correlation function probes the inherent correlations of the ground state, and is nonzero even at equal times $t=0$.  The response function characterizes causality in the system, namely how fast information propagates from a given point to another, and is constrained by the Lieb-Robinson bound to decay exponentially outside of a linear light cone for short-range interactions \cite{lieb72}.

{\it Fermion model.---}In the fermionic model studied in Sec.~\ref{Sec: Fermions}, we will deal with spinless fermions described by annihilation and creation operators, $c_i$ and $c_i^\dagger$, respectively.
In this case too, we are interested in two-point functions that characterize the correlations and the causal response of the model. With two (annihilation and creation) operators at our disposal, the correlation function becomes a $2\times 2$ matrix defined as
\begin{equation}\label{Eq. G 11 fermion}
  i\left[{\mathbf G}_{\s}(t,i-j)\right]_{\alpha\beta}=
   \left\langle {\cal T} c_{i\alpha}(t)  c_{j\beta}^\dagger(0) \right\rangle,
\end{equation}
where $\alpha, \beta\in \{1,2\}$ with $\left(c_{i1} \,\, c_{i2}\right )=(c_i \,\, c_i^\dagger)$ and the fermionic operators given in the Heisenberg picture---the long-ranged ($1/|i-j|^{1+\sigma}$) lattice Hamiltonian is specified in Sec.~\ref{Sec: Fermions}. $\cal T$ is the time-ordering operator defined as ${\cal T}O'(t)O(0)=O'(t)O(0)$ and ${\cal T}O(0)O'(t)=-O'(t)O(0)$ for $t>0$ and fermionic operators $O$ and $O'$ (site indices suppressed). Similarly, the response function is defined as
\begin{equation}\label{Eq. D 11 fermion}
   i\left[{\mathbf D}_{\s}(t,i-j)\right]_{\alpha\beta}= \Theta(t)\left\langle \,[c_{i\alpha}(t), c^\dagger_{j\beta}(0)]_+ \!\right\rangle,
\end{equation}
where the brackets with the subscript $+$ denote the anticommutator. Here too, a Lieb-Robinson bound dictates that the response function is exponentially suppressed outside a linear light cone for short-range interactions \cite{lieb72}.

\

\section{Transverse-Field Ising Model: Scalar Field}\label{Sec: Ising Model}
In this section, we consider the critical and causal properties of the transverse-field Ising model with long-range interactions.
Let us first consider the TFIM with nearest-neighbor interactions in one dimension,
\begin{equation}\label{Eq. TFIM Short-range}
  H=-\sum_{i} \left(\sz_i \sz_{i+1}+ g \, \sx_i \right),
\end{equation}
with $S^{x,y,z}$ the Pauli operators.
This model undergoes a quantum phase transition at $g=1$ from an ordered phase ($g<1$), where the ${\mathbb Z}_2$ symmetry of the model is broken and $\langle \sz_i \rangle \ne 0$, to a disordered phase ($g>1$) where $\langle \sz_i \rangle = 0$. Near the critical point, the long-wavelength behavior of the model can be described by a field theory in the continuum \cite{SachdevBook}, with the Euclidean action
\begin{equation}\label{Eq. Action SR}
    I=\int d\tau \int d x \,\,(\partial_\tau \phi)^2+ (\partial_x\phi)^2+\varrho \phi^2 + u \phi^4\,.
\end{equation}
Here we have rescaled time and spatial coordinates to normalize the coefficients of the derivative terms, defined $\varrho\sim g-1$ characterizing the vicinity to the critical point, and $u$ as the interaction strength. The above action defines the $\phi^4$ field theory in two-dimensional Euclidean space, with $\phi$ a coarse-grained field representing $\langle\sz\rangle$; both are the Ising order parameters which measure the broken symmetry across the phase transition. Indeed, one can see that the scaling dimension of $\sz$, i.e. how its correlations behave under rescaling of space and time coordinates, is consistent with that of the Wilson-Fisher fixed point in two dimensions and with $N=1$ component \cite{SachdevBook}.

Next, we consider the transverse-field Ising model with long-range power-law interactions,
\begin{equation}\label{Eq. TFIM Long-range}
 H_{\sigma}= -\sum_{i \neq j}\frac{\sz_i \sz_j }{|i-j|^{1+\s}}-g\sum_i \sx_i\,,
\end{equation}
where the exponent is defined in terms of $\sigma>0$; the latter is assumed to ensure a well-behaved thermodynamic limit, in which the energy remains extensive as the system size approaches infinity. (The anti-ferromagnetic version of this model, studied in Ref.~\cite{Koffel12}, is not identical to the ferromagnetic model owing to the long-range interactions).  A numerical study of the phase diagram of the above model is performed in Ref.~\cite{knap13}. We also note that long-range interacting {\it classical} Ising models (without the transverse magnetic field but at finite temperature) have also been studied extensively in one dimension \cite{Dyson69,Thouless69}. The model in Eq.~(\ref{Eq. TFIM Long-range}) undergoes a quantum phase transition similar to the short-range TFIM, but at a $\sigma$-dependent critical coupling strength $g$,
whose precise value will not be important for our purposes; we are rather interested in the universal properties of this model, namely the scaling of various correlation functions with distance and time, while the precise value of the critical $g$ can only affect constants of proportionality.
The corresponding universality class {also} includes
models with {additional} ferromagnetic nearest-neighbor, next-nearest-neighbor, or higher but
finite range terms, {added} to the Hamiltonian in Eq.~(\ref{Eq. TFIM Long-range}).

Mapping onto the continuum generates long-range interactions in $\phi$ and its powers provided the original symmetries of the Hamiltonian are respected. For example, $\phi(\tau,x) \phi(\tau,y)/|x-y|^{1+\s}$ and $\phi(\tau,x) \phi^3(\tau, y)/|x-y|^{1+\s}$ are odd with respect to $\phi(\tau,x)\to -\phi(\tau,x)$, i.e.\ a spin flip $\sz_i \to -\sz_i$, similar to the first term in $H_{\sigma}$, but $\phi(\tau,x) \phi^2(\tau,y)/|x-y|^{1+\s}$ is not. For $\s>0$, all long-range terms beyond the quadratic order in $\phi$ and all terms involving spatial derivatives beyond the (local) quadratic gradient term are irrelevant in the sense of RG; additional insertions of the field will make the corresponding term less relevant in the sense of RG, hence $\phi(x)\phi(y)^3/|x-y|^{1+\s}$ is less relevant than $\phi(x)\phi(y)/|x-y|^{1+\s}$, for example. The full action then includes a quadratic piece, containing the long-range term $\phi(\tau,x) \phi(\tau,y)/|x-y|^{1+\s}$, plus the local $\phi^4$ interaction. The former can be more conveniently cast in Fourier space (in imaginary time) \cite{Amit01}, such that
\begin{align}\label{Eq. full action}
I&=I^{(2)}+\int d\tau\int dx~u\phi^4,\\
\label{Eq. quad action--boson}
  I^{(2)}&=\! \int\! d\omega\!\int\! dq \,\,(\omega^2+\varrho+ q^2 + B_\s|q|^\sigma ) \,|\phi(\omega, q)|^2\,.
\end{align}
The parameter $\varrho$ now defines the distance from the critical point of Eq.~(\ref{Eq. TFIM Long-range}); here, we will restrict ourselves to $\varrho\ge 0$, that is, we consider either the critical point, or the paramagnetic side of the Ising transition. The dispersion relation at the quadratic order is simply
\(
  \omega_\s(q)=\sqrt{\varrho+ q^2 + B_\s|q|^\sigma}\,.
\)
Note that the dispersion relation computed by Fourier transforming the coupling constants of the Ising term in $H_{\sigma}$ is more complicated (leading to an expression in terms of polylogarithms).  Here, in the interest of describing the long-wavelength physics that the continuum description is suited to, we have only kept the leading low-$q$ analytical ($q^2$) and non-analytical ($|q|^\sigma$) terms (non-analytical terms with a power smaller than $\sigma$ cannot appear since, upon the inverse Fourier transform, they would give rise to power-laws that decay slower than the original power-law interaction $\sim 1/r^{1+\sigma}$); both will play a crucial role in the following sections. Higher orders of momenta would create finer, or faster-decaying, features, and are thus ignored.

The two-point functions in Eqs.~(\ref{Eq. G 11}) and (\ref{Eq. D 11}) can be described in the continuum as
\begin{align}
 iG_\s(t, x)&\propto\langle {\cal T}\hat \phi(t,x)\hat \phi(0, 0)\rangle,\\
 iD_\s(t,x)&\propto\Theta(t) \langle [\hat \phi(t,x), \hat \phi(0,0)]\rangle,
\end{align}
where $\hat \phi$ denotes the operator-valued field in the Heisenberg picture. In these equations, we have only given the proportionality relations, as we are ultimately interested in scaling relations but not the precise coefficients---the coarse-graining of the lattice operators to their continuum counterparts will introduce nonuniversal lattice-constant-dependent coefficients, while the long-distance/time behavior are unaffected apart from overall coefficients.
We note that, although Lieb-Robinson bounds do not apply directly to the continuum model being studied, sufficiently close to criticality the continuum response function gives a quantitative description of the lattice response function on length scales much larger than the lattice spacing, and hence encodes the correct asymptotic scaling with space and time of the response function for the underlying lattice model.

In the remainder of this section, we study the continuum description of the Ising model with long-range interactions at or away from criticality within the RG approach. In all cases, we restrict ourselves to the vicinity of the critical point, where the correlation length is large compared to the lattice spacing, and thus a continuum description and the RG treatment are well justified.

\subsection{Quadratic model}
To gain some insight into the causal structure and correlations of the long-range interacting model, we first study the quadratic part of the action, but consider the effects of the interaction via RG in the next section.
At the quadratic order, the two-point correlators above completely characterize the system, and can be directly extracted from the dispersion relation as (see App. \ref{App. 2pt functions})
\begin{equation}\label{Eq. corr fn}
     G_\s(R)\propto\int_0^\infty \frac{dq}{\omega_\s(q)}\,\cos(q R)\,
\end{equation}
for the correlations at equal times [$G_\s(R)\equiv  G_\s(0,R)$] and
\begin{equation}\label{Eq. res fn}
   D_\s(t, R)\propto \int_0^\infty \frac{dq }{\omega_\s(q)} \, \sin(\omega_\s(q) t)\cos(q R)\,
\end{equation}
for the response function.  Here we denote the distance by $R=|x|$, and have only given the proportionality relations, as we are ultimately interested in scaling relations and not the precise coefficients.

We start by making a simple observation that, for $\sigma<2$, the long-range term ($|q|^\sigma$) appears to be more relevant than $q^2$, in which case one can simply drop the latter from Eq.~(\ref{Eq. quad action--boson}). However, a more careful treatment reveals that, at least at criticality, the long-range interaction, considered as a perturbation on top of the short-range interacting model in Eq.~(\ref{Eq. Action SR}), can be dropped for values of $\sigma> 2-\eta_{\rm SR}=7/4$ where $\eta_{\rm SR}=1/4$ is the anomalous dimension of the field $\phi$ in the short-range interacting model (\ref{Eq. Action SR}).
In the remainder of this subsection (at the level of the quadratic action), we shall ignore this complication, but discuss it in some detail in
Sec.~\ref{Sec. RG}.
We also remark that a specific feature of the model in Eq.~(\ref{Eq. TFIM Long-range}) is that $B_\s>0$ for $0<\s<2$, and $B_\s<0$ for $2<\s<4$, a pattern repeated periodically; this can be easily seen by computing the Fourier transform of the $1/|i-j|^{1+\s}$ power law in Eq.~(\ref{Eq. TFIM Long-range}).
The fact that $B_\s$ is positive when it is relevant is indeed assuring as the Hamiltonian is bounded from below. On the other hand, when the corresponding term is not relevant, we may have $B_\s<0$ which does not pose a problem as one must impose a high-momentum cutoff naturally provided in lattice systems. Below, we study these two cases separately.

For the sake of comparison, we first quote the two-point functions in the absence of long-range interactions at or away from criticality. First, the correlation function for the quadratic model at criticality is given by $G_{\rm SR}(R)\sim \log R$, while the full (short-range) interacting model yields $G_{\rm SR}(R) \sim 1/R^{\eta_{\rm SR}}$ \cite{SachdevBook}. Away from criticality, the correlation function decays exponentially beyond a length scale defined as the correlation length $\xi$, i.e., $G_{\rm SR}(R)\sim \exp(-R/\xi)$.  The response function is given by
\begin{equation}\label{Eq. D short-range}
  D_{\rm SR}(t, R) \sim
  \begin{cases}
  \Theta(t-R), & \varrho=0\,,  \\
  \Theta(t) \im\left[K_0\left(\sqrt{\varrho}\sqrt{R^2-t^2}\right)\right], & \varrho> 0\,,
  \end{cases}
\end{equation}
that is,
it vanishes identically outside the linear light cone $t=R$, and, for $\varrho=0$, it is simply a constant within the light cone. We note that the {response functions in Eq.\ (\ref{Eq. D short-range})} would be modified if interactions were included, and thus serve only as approximations to the response functions for the short-range interacting model.

\subsubsection{$0<\sigma<2$ at criticality}\label{Sec. sigma<2 at criticality}
At the critical point, the dispersion relation can be approximated as
   \(
    \omega_\s(q) \approx  |q|^{\s/2},
   \)
   where we have set $B_\s=1$. The correlation function can be computed from Eq.~(\ref{Eq. corr fn}) via a simple rescaling of $q$ by $1/R$, yielding
  \begin{align}\label{Eq. corr fn criticality}
    G_\s(R)
        &\sim \frac{1}{R^{1-\sigma/2}}\,.
  \end{align}
  The exponent of this power-law defines the scaling dimension
  of the field at the quadratic order; however, it will receive corrections in the course of RG, as we shall discuss in Sec.~\ref{Sec. RG}.

{
  The response function is obtained from Eq.~(\ref{Eq. res fn}) via a similar rescaling, yielding
  \begin{align}\label{Eq. fn g}
    D_\s(t,R)
         \sim &\frac{1}{R^{1-\sigma/2}} \int_0^\infty \frac{dq}{q^{\sigma/2}} \,\, \sin\left[q^{\sigma/2}\frac{t}{R^{\sigma/2}}\right]\cos q  \nonumber \\
         =& \frac{1}{R^{1-\sigma/2}} \,\,  g_\sigma\!\!\left[\frac{t}{R^{\sigma/2}}\right]\,,
  \end{align}
  where we have defined the scaling function $g_\sigma$ in the last equality.
  Note that, by extending the momentum integration to infinity, we have derived a simple scaling relation for $D_\s(t,R)$. We point out that a similar scaling relation also emerges for the time-ordered correlation function $G_\s(t,R)$, though with a different functional form.
  Nevertheless, one should keep in mind that the above expression applies only to regions well outside the linear light cone $R= t$ since we have dropped the momentum cutoff as well as the high-momentum modes (including $q^2$)
  from the dispersion relation. We also note that the multiplicative power-law in Eq.~(\ref{Eq. fn g}) is identical to Eq.~(\ref{Eq. corr fn criticality}) simply due to the scaling dimension of the field $\phi$; apart from this power-law required for dimensional reasons, the information about the causal behavior is encoded in the scaling function $g_\s$.
  For small arguments, $g_\s(s)\sim s^3$ independent of $\s$, which results in $D_\s(t,R)\sim t^3/R^{1+\s}$. Thus, at fixed $t$, the response function decays as $1/R^{1+\s}$ with distance, consistent with the demands of the Hastings-Koma bound \cite{hastings05} applied to $H_{\sigma}$ \footnote{Another way to see this is to start from Eq.~(\ref{Eq. fn g}) and make the analytic continuation to the imaginary axis, $q\to i q$, where the response function becomes
  $D_\s(t,R)\sim \im \int \frac{dq}{\omega_{iq}}\sin(\omega_{iq} t) e^{-q R}$.
  Since $\omega_{iq }\sim q^{\s/2} < q$ for large $q$, the exponential growth of the sine function is countered by the exponential decay in $q$. For large $R$, only a small region in $q\ll 1/R$ contributes, which makes it possible to expand the sine function. One can then see that the expansion of the sine function to the third order gives the leading-order non-local contribution in the text.}.
  The function $g_\sigma(s)$ increases monotonically with $s$ up to $s\sim 1$, but, beyond this point, exhibits an oscillatory behavior to be further discussed below.
  Hence, the response function at a distance $R$ away from a local quench at $t=0$  reaches the first local maximum in time around
  \begin{equation}\label{Eq. z quadratic order}
  t\sim R^{\sigma/2}\,.
  \end{equation}
  \begin{figure}[t]
    \centering
    \includegraphics[width=8cm]{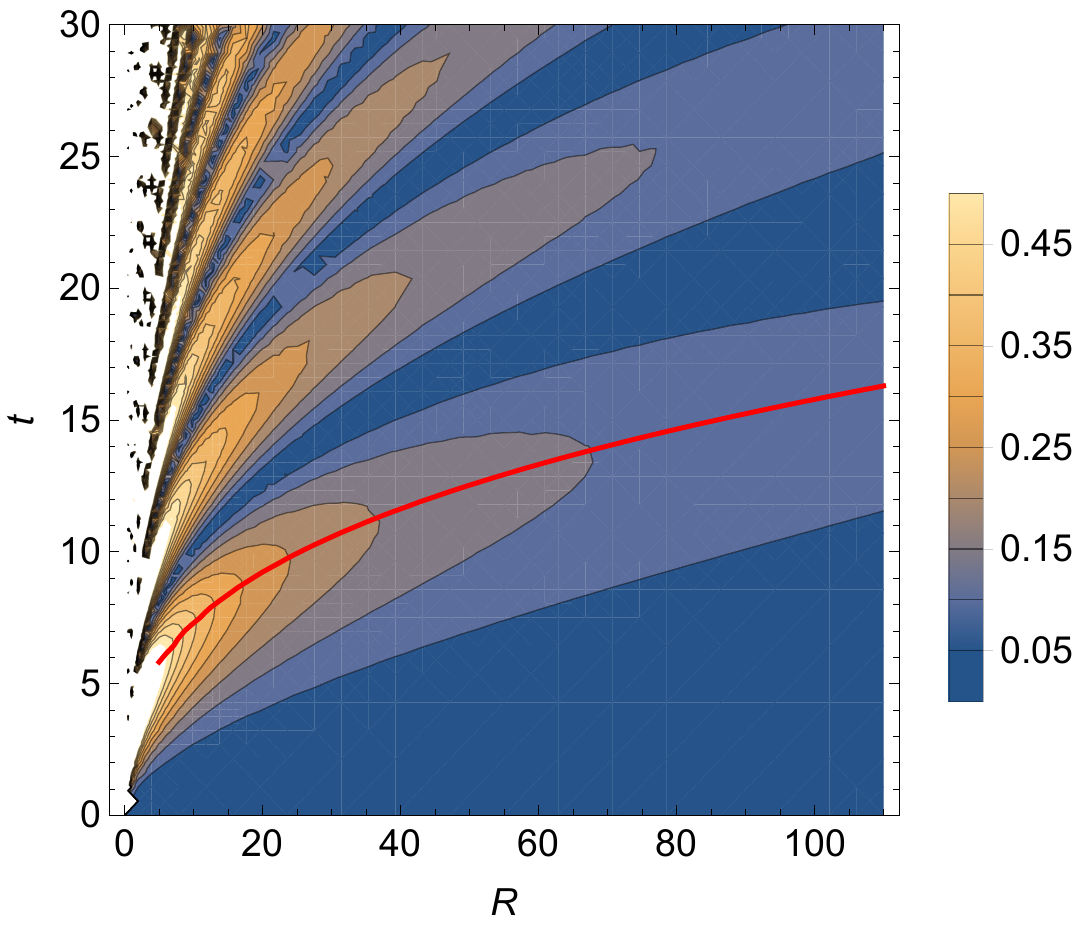}
    \caption{Contour plot of the (unnormalized) response function $D_{2/3}(t,R)$. The response function is computed via the scaling function $g_{2/3}$. The causal region exhibits a nonlinear behavior with a finger-like pattern extending beyond the linear light-cone structure of short-range interactions. The numerical evaluation of the response function with the lattice dispersion relation and momentum cutoff
    produces an almost identical plot.
    The (red) solid line represents the curve $t\sim R^{\s/2}$ with $\s=2/3$.}%
    \label{Fig. Nonlinear}%
  \end{figure}
This implies that the causal region obeys a power law, in sharp contrast with short-range interactions. A numerical evaluation (at the quadratic order) with the full dispersion relation shows that the response function branches outside the {linear} light cone polynomially as $t\sim R^{\sigma/2}$; see Fig.\ \ref{Fig. Nonlinear}. In fact, by setting $D_\s(t,R)$ to a small constant, we find polynomial \emph{fingers} reaching outside the linear light cone $t=R$, which are, however, finite in extent as the signal weakens while it propagates. In general, the effect of a local quench decays at long times, and thus contours of the response function---due to a local perturbation---are typically finite in extent. The fact that the response function for short-range interactions [Eq.~(\ref{Eq. D short-range})] does not decay at long times is an artificial feature of the quadratic action $I^{(2)}$, and would not be the case in the presence of interactions, in higher dimensions, or sufficiently far from the critical point (where RG irrelevant terms that break relativistic invariance must be included).  Therefore, to correctly identify the causal behavior, we
disregard the decaying power-law function out in front (in this case, $1/R^{1-\s/2}$), and focus on the scaling function $g_\s$.

  The power law in Eq.~(\ref{Eq. z quadratic order}) identifies the dynamic critical exponent that characterizes the relative scaling of time with respect to space. For $\sigma>2/3$, RG calculations produce corrections to the mean field value (i.e. calculated at the quadratic order) of the exponent above, as we shall discuss in Sec.~\ref{Sec. RG}.
Scaling relations of the form (\ref{Eq. fn g}) are generic beyond the quadratic model.  However, within the quadratic model,
an explicit form for the scaling function $g_\sigma$ can be obtained:
  For all $0<\s<2$ it undergoes periodic oscillations in units of of $ s^{1/(1-\s/2)}$, with an asymptotic envelope function
  \begin{equation}
      \amp[g_\sigma(s)]\sim s^{(1-\s)/(2-\s)}\,.
  \end{equation}
For example, for $\s=1$, we have $g(s)\sim \cos(s^2/4)+{\cal O}(s^{-1})$. Compared with numerics, the scaling function $g_\s$ is quantitatively accurate for the low lying fingers, but also captures the qualitative features of higher ones.

\subsubsection{$\sigma>2$ at criticality}
In this regime, long-range interactions are not relevant, and one thus expects to recover the same asymptotic form of the correlation function as well as the linear light-cone structure of short-range interactions.
However, the slight deviation from a linear light cone can be quantified, and shown to take a universal form.
For convenience, we restrict ourselves to $2<\s<4$, since in this range higher-order analytical terms in momentum ($q^4, q^6, \cdots$) can be dropped compared to $|q|^\s$.
The physics at long times and distances is dominated by its behavior at low momentum, where the dispersion relation can be approximated as
\(
  \omega_\s(q)\approx |q| - |q|^{\s-1}.
\)
Here we have dropped the coefficient of the non-analytic term, but kept the correct (negative) sign inherited from $B_{\sigma}$.
We also approximate the frequency in the denominator of Eq.~(\ref{Eq. corr fn}) by $\omega_\s(q)\approx q$. This set of approximations is valid in the vicinity of the linear light cone $t=R$, which is the main focus here. For space-time points well inside or outside the light cone, the response function is generally not captured by the following scaling relations.
With the above approximations, the response function can be recast as
\begin{align}\label{Eq. response function 1d}
    D_\s(t,R)\sim &\int_0^\infty \frac{dq}{q} \,\, {\Big\{}\sin[q(t-R)-q^{\sigma-1}t] \nonumber \\
    &\hskip .5in +\sin[q(t+R)-q^{\sigma-1}t]{\Big \}} \nonumber \\
    =& \tilde f_\s\left[\frac{t-R}{t^{{1}/({\sigma-1})}}\right]+\tilde f_\s\left[\frac{t+R}{t^{{1}/({\sigma-1})}}\right]\,. \nonumber
  \end{align}
  At sufficiently long times ($t\gg 1$), the argument of the last term is large since $t^{1/(\sigma-1)} \ll t$ for $\sigma>2$. In this limit, one can see that $\tilde f_\s(s) \to {\rm const}$ as $s\to +\infty$. Therefore, the response function can be written as
  \begin{equation}\label{Eq. Scaling eq}
    D_\s(t,R) \sim f_\sigma\left[\frac{t-R}{t^{{1}/({\sigma-1})}}\right],
  \end{equation}
  where the function $f_\s(s)\equiv \tilde f_\s(s)+\tilde f_\s(+\infty)$ has the properties
  \begin{equation}
    f_\s(s)=
    \begin{cases}
      {\rm const}, & s \to +\infty\,, \\
      0, & s \to -\infty\,.
    \end{cases}
  \end{equation}
  This implies that the response function approaches a constant (zero) inside (outside) the linear light cone.
  \begin{figure}[t]
    \centering
    \includegraphics[width=8cm]{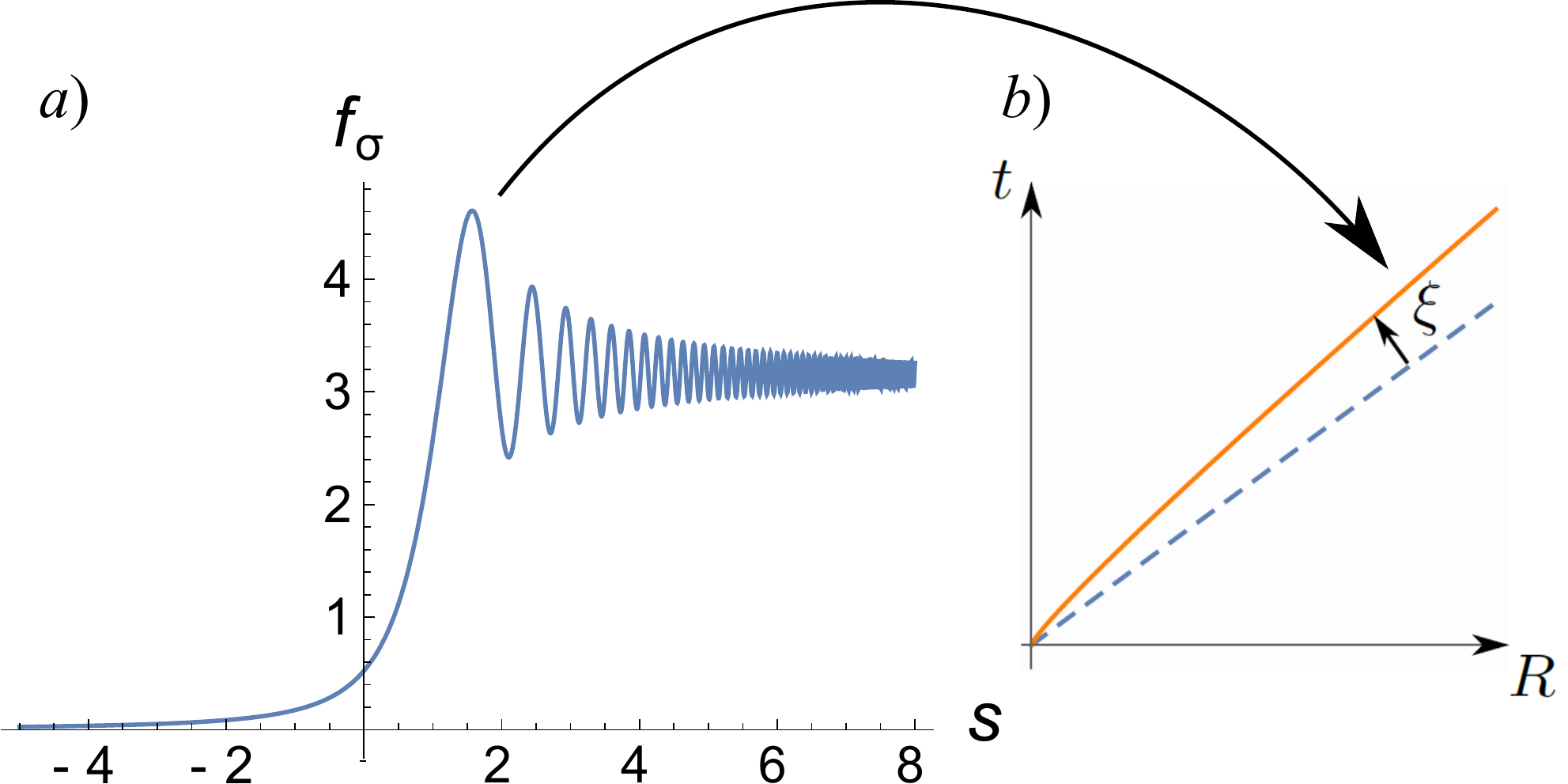}
    \caption{(a) The scaling function $f_\s$ characterizing the response function as a function of $s=(t-R)/t^{1/(1-\sigma)}$ for $\s=5/2$. The function $f_\s$ approaches a constant inside the linear light cone ($s>0$), but decays rapidly outside the light cone ($s<0$). (b) The first local maximum of the response function (solid curve) is bent inside the light cone (dashed line). $\xi$ characterizes the deviation from the linear light cone. For long times, $\xi/t\to0$, and the causal region becomes a sharp linear cone.}%
    \label{Fig. Linear}%
\end{figure}
  The crossover between the two limits only depends on $R$ and $t$ in the particular combination in Eq.~(\ref{Eq. Scaling eq}). To find the causal region, we set the response function to a (small) constant which then yields a constant value for the argument of the function $f_\s$,
  \begin{equation}\nonumber
      \frac{t-R}{t^{{1}/({\sigma-1})}} = \mbox{(large) const}.
  \end{equation}
  For sufficiently long times, {we have}
  \(
      t^{1/(\sigma-1)} \ll t-R
  \)
  near $R\approx t$, which yields a nearly-linear light cone; however, the causal region is not a sharp cone as it spreads out over a region of size $t^{1/(\sigma-1)}\ll t$. The deviation of the causal region from the linear light cone can be characterized by $\xi =t-R$; the above considerations yield
  \begin{equation}
    \frac{\xi}{t} \sim t^{-(\s-2)/(\s-1)}\,.
  \end{equation}
  At long times, $\xi/t\to 0$ for $\s>2$, and the linear light cone becomes exact.
  For the quadratic model considered in this section, one can sometimes (for certain values of $\sigma$)
  obtain an analytical expression for the scaling function in Eq.~(\ref{Eq. Scaling eq}).
  We find that the causal region is slightly bent \emph{inside} the linear light cone [with $\xi(t)\sim t^{1/(\s-1)}$], see Fig.~\ref{Fig. Linear}. The fall-off outside the light cone is monotonic, with asymptotics governed by
  \begin{equation}
    f_\s(s)\sim |s|^{1-\s}\,.
  \end{equation}

  \subsubsection{All $\sigma>0$ away from criticality}\label{Sec. All sigma noncritical}
  We first compute the correlation function $G_\s(R)$ away from criticality for all $\s>0$ not being an even integer. With short-range interactions, this function decays exponentially beyond the correlation length; however, long-range interactions always induce a power-law decay of the correlation function.
 To be completely general
 , we can consider a dispersion relation
 of the form $\omega_\s(q)=\sqrt{1+F(q^2)+B_\s |q|^\sigma}$ \, where $F(q^2)$ is a polynomial function of $q^2$ [with $F(q^2\to 0)=0$] such that the expression under the square root is nonnegative; while we usually truncate the function $F(q^2)$
 at the quadratic order to capture the long-wavelength physics, it generally has a more complicated, but analytic, form. Equation~(\ref{Eq. corr fn}) can be cast as an integral over $q\in (-\infty,\infty)$ with the substitution $\cos(q R) \to e^{iq R}$. One can then deform the contour of integration to the upper-half plane in the complex plane. However, the function $|q|^\sigma$ should be treated separately as an analytic function for $q \gtrless 0$. We thus have to find the zeros, or branch points, of $\omega_\s(q)$ in the upper half-plane. For a value of $\sigma$ that is not an even integer, there are no zeros on the real or imaginary axis. The branch points are denoted by $q_*$, which, with a convenient normalization, can be assumed to be of the order of unity, and specifically $\im q_*\sim 1$. The contribution of a branch cut is exponentially suppressed as
 \begin{align}\label{Eq. Exp decay}
   \chi(R) \, e^{-({\footnotesize\im} q_*) R},
 \end{align}
 up to a multiplicative polynomially decreasing function $\chi(R)$. For large $R$, the branch cut only contributes near $q\sim q_*$, thus $\omega_\s(q) \sim \sqrt{q-q_*}$ and $\chi(R) \sim \int dq \frac{1}{\sqrt{q}} e^{-qR} \sim1/\sqrt{R}$. On the other hand, the integral along the imaginary axis gives
 \begin{align}\label{Eq. corr fn noncrit}
G_\s(R) &\sim  \int_0^{\infty} dq \, e^{-q R} \, \im \frac{1}{\sqrt{1+F(-q^2) +e^{i\pi \sigma/2}q^\sigma}}\nonumber \\
 &\sim \int_0^{\infty} dq \, e^{-q R} \, q^\sigma \sim \frac{1}{R^{1+\sigma}}\,,
\end{align}
up to exponentially decaying corrections in Eq.~(\ref{Eq. Exp decay}). Depending on the relative ratio of the coefficients of the exponential and power-law terms in Eqs.~(\ref{Eq. Exp decay}) and (\ref{Eq. corr fn noncrit}), we may find an exponentially decaying correlation at intermediate length scales followed by a power law at large distances, consistent with Refs.\ \cite{gong14,Pupillo14,Gong15}.
In the last step, we have expanded the denominator around $q=0$ to obtain the large-distance behavior, which yields the exact asymptotic form of the correlation function. Replacing $q^\s$ by an analytic expression such as $q^n$ with $n$ an even integer, this argument breaks down
as $\im (1/\omega_{q=i|q| })=0$ on the imaginary axis.

The exponent characterizing the decay of the correlation function away from criticality [Eq.\ (\ref{Eq. corr fn noncrit})] {thus} coincides with the $1+\sigma$ exponent of the long-range interactions. While this result is derived at the quadratic level of the action, it holds more generally even in the presence of interaction terms, as we shall discuss in the next section.

  Next we study the response function. Away from criticality, scaling relations take a more complicated form, and analytical expressions are less available. Instead we shall resort to a simple analysis of the group velocity $v_q=\partial \omega_\s(q)/\partial q$. A semiclassical approximation suggests that the causal region should be linear when the velocity is bounded from above \cite{Calabrese06}, see also \cite{Rajabpour15}.
  The group velocity at high momenta is typically bounded due to the short-wavelength cutoff; however, for sufficiently long-range interactions, it may diverge as $q\to 0$. At criticality, the semiclassical picture correctly predicts a linear causal behavior for $\s>2$. Away from criticality, the dispersion relation is $\omega_\s(q)=\sqrt{1+q^2+B_\sigma |q|^\sigma}$, where we have set $\varrho=1$.
  An inspection of the group velocity then shows that, in the limit where $q\to 0$, it diverges for $\s<1$, but approaches a constant (or zero) for $\s\ge 1$.
  For the special value of $\s=1$ at the borderline between the linear and non-linear causal behaviors, the dispersion relation is approximately $\omega_1(q)\approx \sqrt{1+2|q|}$, where we have set $B_\s=2$ for notational convenience.
\begin{figure}[t]
    \includegraphics[width=8cm]{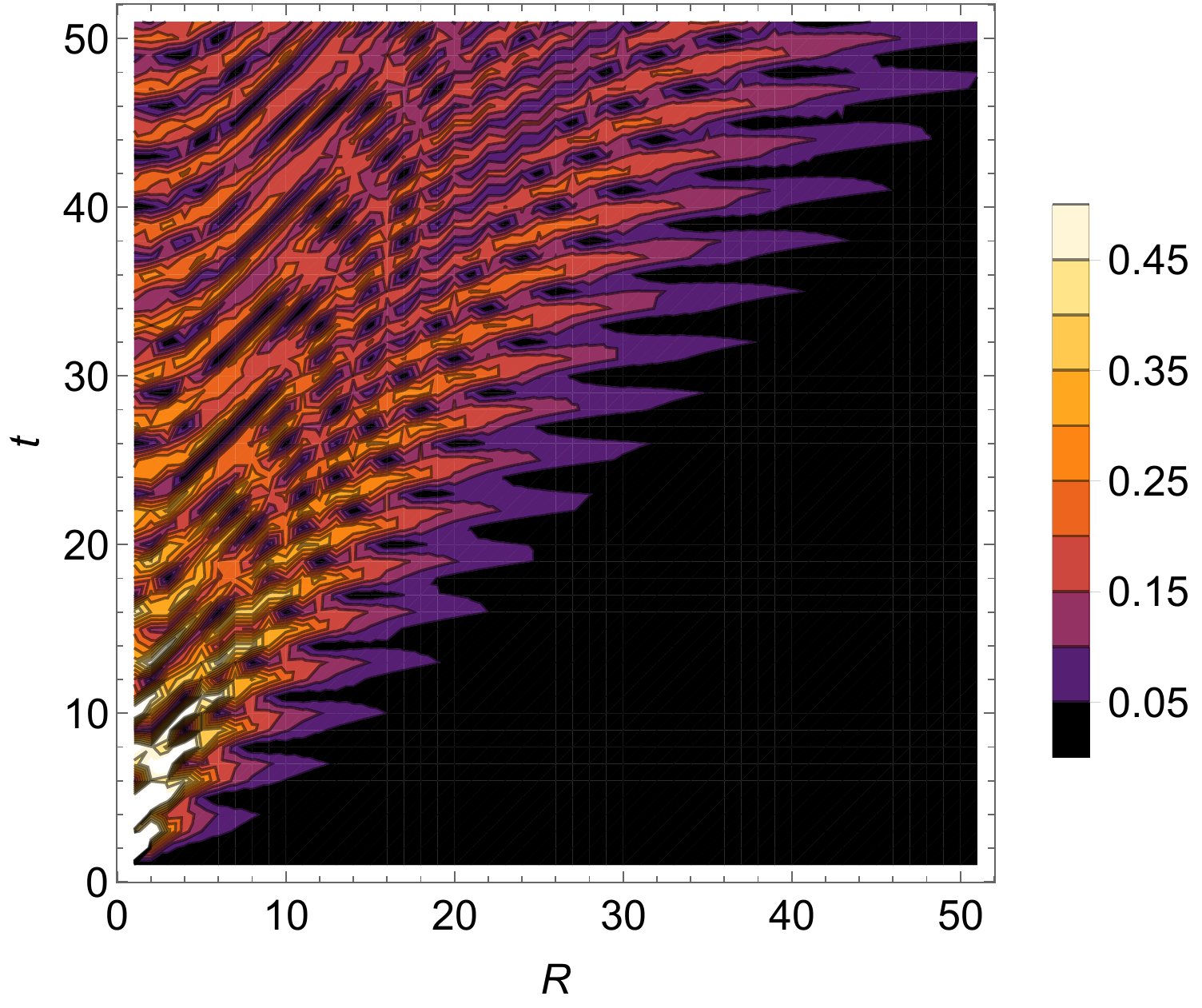}
    \caption{The norm of the (unnormalized) response function $|D_1(t,R)|$ away from criticality evaluated with the full dispersion relation $\omega_1(q)= \sqrt{1+2q-q^2/\pi}$  for $q\in (0,\pi)$ (with $\varrho=1$ and the $q$-dependence obtained from the polylog function of the lattice dispersion relation at $\sigma=1$). The analytical expression in the text produces a very similar plot sufficiently away from the vertical line ($R=0$). The response function exhibits a linear causal behavior. }
    \label{Fig. Noncritical1}%
\end{figure}
  In this case, one can find an exact analytical expression for the response function as
  \begin{align}\label{Eq. Fersnel}
  D_1(t&,R)\sim  {\rm Re}{\Bigg \{}\frac{e^{\frac{i\pi}{4}}}{  \sqrt{R}} e^{-i(R^2+t^2)/2R} \nonumber \\
  &\times \left[\text{erf}\left(e^{\frac{i\pi}{4}}\,\frac{ R-t}{ \sqrt{2R}}\right)-(t\to -t)\right]{\Bigg \}},
  \end{align}
  where $\mbox{erf}$ is the error function.
  The expression inside the curly brackets falls sharply beyond a straight line $t=R$, which thus indicates a linear boundary of the causal region. The response function, i.e., the real part of the expression in curly brackets, is highly oscillatory, but exhibits the same causal behavior as the expression inside the brackets.
  Furthermore, the amplitude of the response function
  decays as $1/R^2$ for fixed $t$ outside the linear light cone, consistent with the Hastings-Koma bound \cite{hastings05} for the lattice model (\ref{Eq. TFIM Long-range}). The response function computed with the full dispersion relation, plotted in Fig. \ref{Fig. Noncritical1}, clearly exhibits a causal region with a linear boundary. We also note that response functions are generically highly oscillatory away from criticality even for the short-range interacting model [Eq.~(\ref{Eq. D short-range})].

  Next we consider {$\sigma < 1$.}
  In the absence of analytical expressions, it is more difficult to fully characterize the causal behavior. Naively looking at the response function could be misleading as setting $D_\s(t,R)=$const gives rise to a vertical line at long times in some cases, and does not clearly identify the causal behavior. Instead we set $t\sim R^\alpha$ in $D_\s(t,R)$ and plot it as a function of $R$. We can then identify $z$ as the value of $\alpha$ for which the response function obtains a simple power-law form.  The function $D_\s(R^\alpha,R)$ is plotted in Fig.~\ref{Fig. ThreeAlpha} for $\sigma=1/3$ and three different values of $\alpha$. Clearly, only for $\alpha=\sigma/2=1/6$ does this function have a particularly simple form, falling off with the distance as a power-law $1/R^{1+\s}$ for $\s=1/3$.  As expected, the associated exponent is identical to the one characterizing the decay of $G_\s(R)$ at long distances in Eq.~(\ref{Eq. corr fn noncrit}). These observations are consistent with the scaling form $D_\s(t,R)\sim R^{-(1+\s)} \, {\mathfrak g}_\s\!\!\left(t/R^{\s/2}\right)$ with $\s=1/3$ and ${\mathfrak g}_\s$ a scaling function, at least for the set of parameters studied here.
  The scaling function ${\mathfrak g}_\s$ defines the relative scaling of space and time coordinates, giving a dynamic critical exponent $z=\s/2$ for values of $0<\s<1$; and in this sense the causal behavior is nonlinear.
  \begin{figure}
   \centering
    \includegraphics[width=9.5cm]{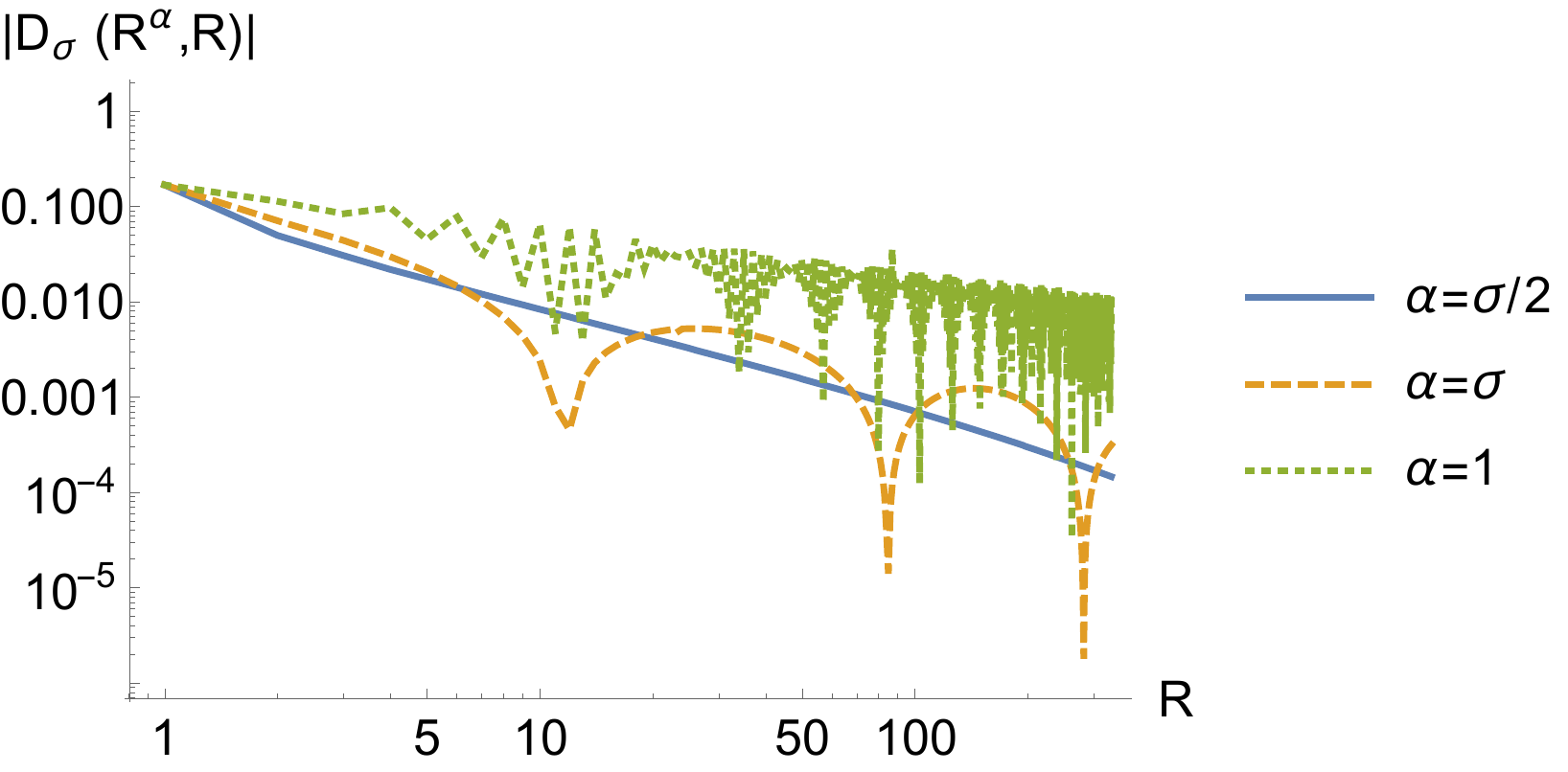}
     \caption{The norm of the (unnormalized) response function $|D_\s(t,R)|$ evaluated at $t = R^\alpha$ for $\s=1/3$ away from criticality. The three plots correspond to different choices of $\alpha=\s/2$, $\s$, and 1 (with $\s=1/3$). The dispersion relation is taken as $\omega_{1/3}(q)=\sqrt{1+\frac{2}{\sqrt{3}}\Gamma\left(\frac{-4}{3}\right)q^{1/3}+\frac{1}{2}\zeta\left(\frac{-2}{3}\right)q^2}$    for $q\in (0,\pi)$ (with $\varrho=1$ and the $q$-dependence obtained from the polylog function of the lattice dispersion relation at $\s=1/3$). For the exponent $\alpha=\s/2$, we find a simple power-law decay of the response function with $R$, suggesting $z=\sigma/2$.}
     \label{Fig. ThreeAlpha}
   \end{figure}
  This is identical to the dynamic critical exponent of Sec.~\ref{Sec. sigma<2 at criticality} computed for $0<\s<2$ at criticality, while, in the noncritical case studied in this section, the above identification is valid only for $0<\s<1$. We stress that this conclusion is based on our numerical results for a limited range of parameters; an extensive investigation of this regime should {be carried out to} confirm
  its universal scaling properties.

\subsection{Beyond the quadratic model: A renormalization-group study}\label{Sec. RG}
To study the full interacting model in Eq.\,(\ref{Eq. full action}), we first recall that long-range interactions are irrelevant for $\s>2-\eta_{\rm SR}=7/4$, in which case the renormalization group and the universal properties and exponents at criticality are those of the usual $\phi^4$ model in one higher dimension than the original model. The borderline value of $\s=2-\eta_{\rm SR}$ arises due to the following consideration. We can treat the long-range interaction $|q|^\s |\phi(\omega,q)|^2$ as a perturbation on top of the short-range interacting model in Eq.~(\ref{Eq. Action SR}), for which the scaling dimension of the field is $[\phi]=\eta_{\rm SR}/2$. One can then easily see that the scaling dimension of the long-range interacting {term} becomes negative for values of $\s>2-\eta_{\rm SR}$. This feature also arises in classical long-range interacting systems \cite{Sak73}, see also \cite{Cardy96}, and also \cite{Federico14} for a recent review.
We thus first write the action for $0<\s<2-\eta_{\rm SR}$ in real space and (imaginary) time in $d$ spatial dimensions as
   \begin{align}
    &I=\int \!\!d\tau \!\!\int\!\! d^d\bx \,\, \left[A\, (\partial_\tau \phi)^2+\varrho\phi^2 \right]  \nonumber \\
    &+B_\sigma \int \!\!d\tau\!\!\int \!\! d^d\bx \!\! \int \!\!d^d \by\frac{\phi(\tau,\bx)\phi(\tau,\by)}{|\bx-\by|^{d+\sigma}}+u\int \!\!d\tau \!\!\int\!\! d^d\bx \,\,\phi^4\,, \nonumber
   \end{align}
   with the (possibly normalized) coefficients $A$, $B_\s$, and $\varrho$.
   As a first step, we rescale the coordinates and the field as
   \begin{align}\nonumber
     \bx \to \bx'=\bx/b, \quad \tau \to \tau'=\tau/b^z, \quad \phi\to \phi'=b^a \phi\,,
   \end{align}
   where $b$ is an arbitrary constant greater than unity; note that the time and spatial coordinates are scaled differently. The dynamic critical exponent, $z$, and the scaling dimension of the field, $a$, are to be determined.
  Simple dimensional analysis yields the scaling dimension of various terms in the action as
   \begin{align}\label{Eq. Engineered scaling dim}
     &[A]=-z+d-2a, \quad [B_\sigma]=z-\sigma+d-2a, \nonumber \\
     &[\varrho]=z+d-2a, \quad [u]=z+d-4a\,.
   \end{align}
   To find the exponents at the level of mean field, we set $[A]=[B_\sigma]=0$, that is, we require the quadratic part of the action (with the exception of the `mass term' $\varrho$) to be invariant under rescaling. We then obtain
  \begin{equation}\label{Eq. mean-field exponents}
    z=\frac{\s}{2}\,, \qquad \eta=1-\frac{\s}{2}\,,
  \end{equation}
  where we have defined the anomalous dimension $\eta$ of the field via $a=(d-1+\eta)/2$.
  Indeed, with $d=1$, these are the same exponents that describe the two-point functions in Sec.~\ref{Sec. sigma<2 at criticality}. To go beyond mean field, we shall resort to the so-called epsilon expansion. To this end, we first derive the upper critical dimension beyond which the interaction becomes irrelevant, i.e. $[u]< 0$. Demanding that $[u]=0$ at $d=d_u$, one finds $d_u=3\sigma/2$ \cite{Amit01}. Therefore, for our one-dimensional model, the interaction term can be dropped when
  \(
    \s < \frac{2}{3}
  \),
  in which case the field-theoretical arguments indicate that the full model is faithfully represented by the quadratic part, and the mean-field exponents become exact. On the other hand, for $\s\ge 2/3$, these exponents may be modified due to fluctuations, and corrections to them can be obtained as an expansion in $\epsilon=d_u-d = 3\sigma/2-1$ \cite{Wilson72,Wilson74}.

  Before proceeding, we remark that the RG procedure produces only analytical terms, and thus cannot renormalize non-analytical terms in the action. Therefore, the coefficient $B_\s$ in the action is not renormalized, which leads to an exact non-renormalization condition from Eq.~(\ref{Eq. Engineered scaling dim}) by setting $[B_\s]=0$ (cast in terms of the anomalous exponent $\eta$)
  \begin{equation}\label{Eq. exact eta and z}
    \eta=1+z-\sigma\,,
  \end{equation}
  valid to all orders of perturbation theory.  Thus finding one of the two exponents ($\eta$ or $z$) completely determines the other one. (For long-range interacting classical systems, i.e., in the absence of the imaginary-time direction, the exponent $\eta$ assumes the fixed value $\eta=2-\s$ \cite{Fisher72}.)

  The $\epsilon$-expansion can be organized as a series of loop diagrams. The two-point function is renormalized by the loop diagrams depicted in Fig.~\ref{Fig. 2ptLoops} to the two-loop order.
  \begin{figure}[h]
  \centering
  \includegraphics[width=5cm]{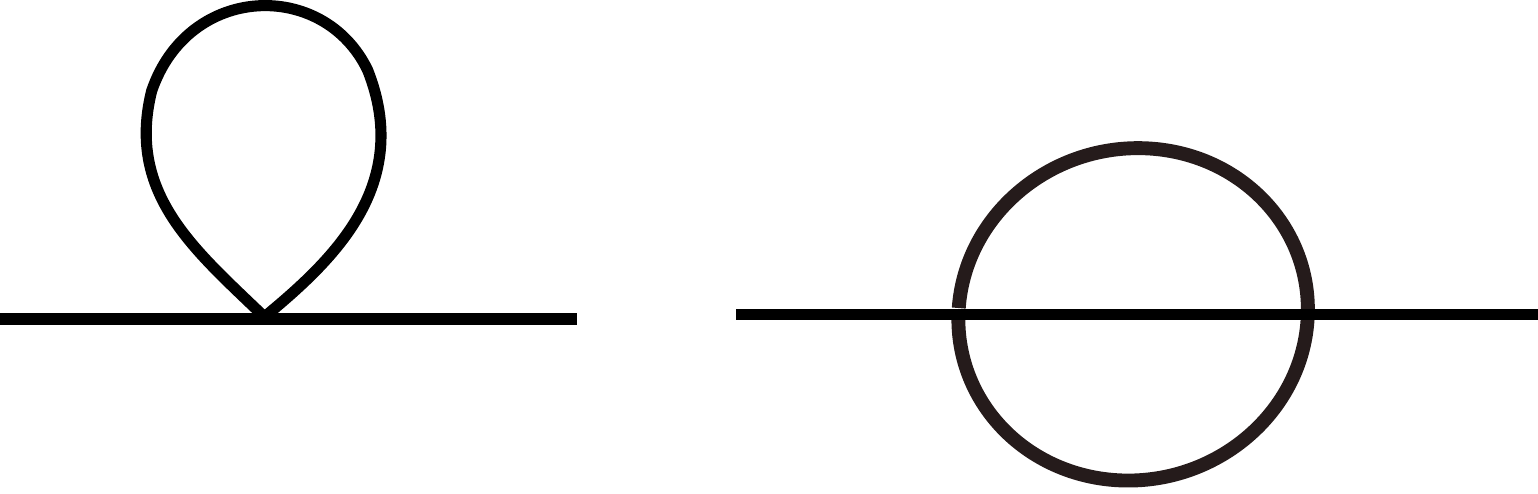}
  \caption{The one- and two-loop diagrams contributing to the two-point function.}\label{Fig. 2ptLoops}
\end{figure}
  However, the one-loop diagram does not contribute to the exponents $\eta$ or $z$ since the loop correction is independent of the (imaginary) frequency and momentum of the external lines. Therefore, we should consider the two-loop diagram, which is significantly harder to compute. We leave the details of RG to App.~\ref{App. RG}, but quote the result of the second-order $\epsilon$-expansion for the exponents $z$ and $\eta$. Defining $\Delta z$ and $\Delta \eta$ as the difference of the exponents $z$ and $\eta$ from their mean-field values in Eq.~(\ref{Eq. mean-field exponents}), we find
  \begin{equation}\label{Eq. Two-loop}
    \Delta \eta= \Delta z= \varsigma(\sigma) \epsilon^2+ {\cal O}(\epsilon^3)\,,
  \end{equation}
  where ${\varsigma}(\s)$ is a rather complicated expression reported in App.~\ref{App. RG}, but is approximately reproduced by ${\varsigma}(\s)\approx 1/[24 (1+\sigma^2)]$. (For the special case of $\sigma=1$, this is consistent with the result obtained in Ref.~\cite{Sachdev04}, see App.~\ref{App. RG}.)
   Computing the critical exponents in $d=1$ dimension, we have $\epsilon=3\sigma/2-1$ as explained above. We briefly note that, while $\epsilon$ can be of order 1, $\epsilon-$expansion has been remarkably successful even for $\epsilon=1, 2$ \cite{KardarBook}.
\begin{figure}[h]
  \centering
  \includegraphics[width=8.5cm]{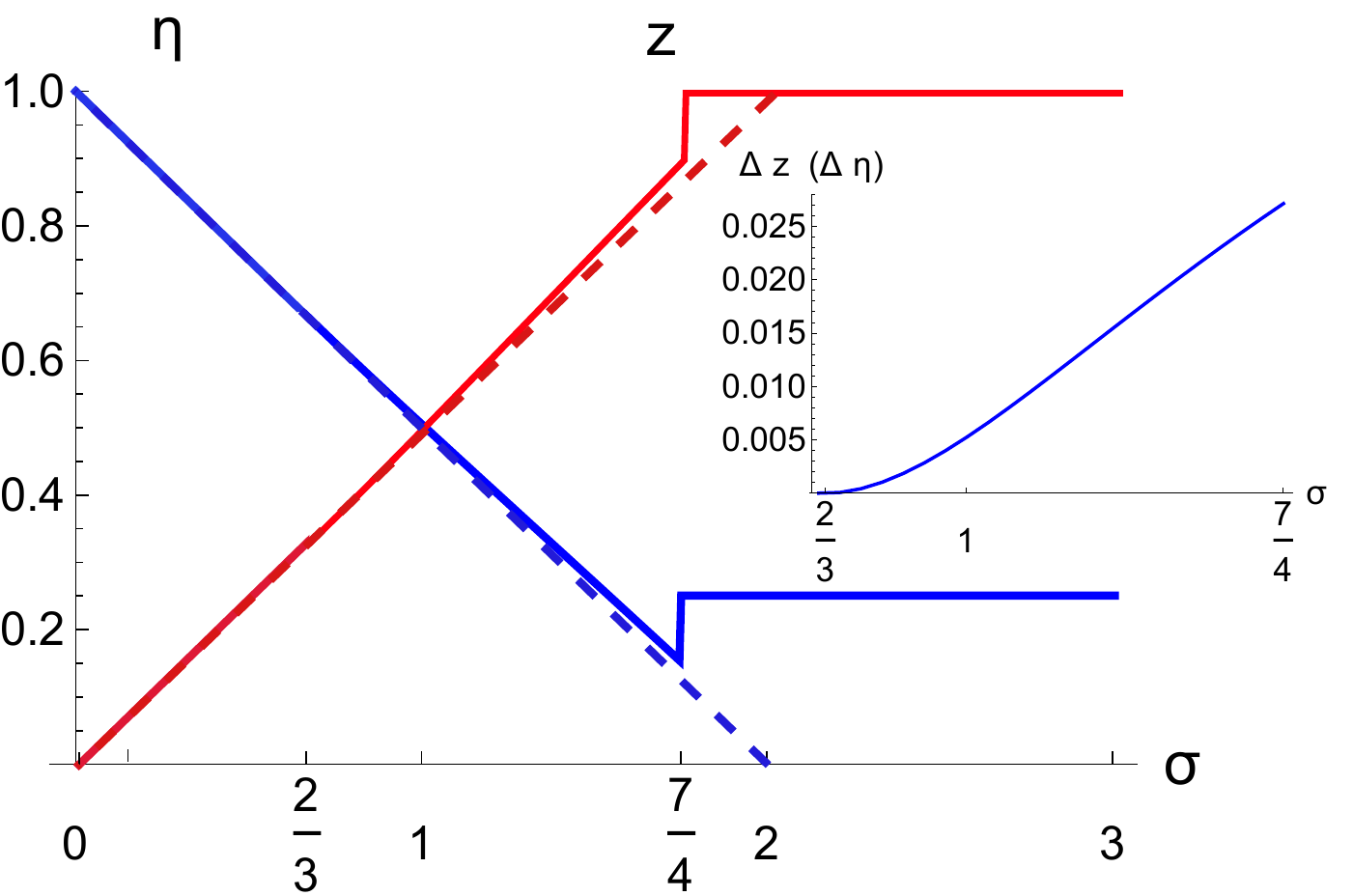}
  \caption{The critical exponents $\eta$ and $z$ as a function of $\s$. These exponents assume their short-range values $\eta=\eta_{\rm SR}=1/4$ and $z=1$ for $\s>7/4$. For $\s<2/3$, they stick to their mean-field values $\eta=1-\s/2$ and $z=\s/2$, and, for $2/3<\s<7/4$, their values are computed up to the two-loop order (the dashed lines show the mean-field exponents). The inset shows $\Delta z=z-\sigma/2$ and $\Delta\eta=\eta-(1-\sigma/2)$, the difference of the critical exponents from their mean-field values, up to the two-loop order.}\label{Fig. ssigma}
\end{figure}
  The exponents $\eta$ and $z$ as a function of $\s$ are plotted in Fig.~\ref{Fig. ssigma}. For $\sigma<2/3$, field theory indicates that these exponents stick to their mean-field values, while for $\sigma>7/4$, they assume their short-range values. For intermediate values of $2/3<\s<7/4$, the two-loop correction gives a slight deviation from the mean-field prediction (the latter shown as the dashed line in the range $2/3<\s<7/4$ in Fig.~\ref{Fig. ssigma}). The two-loop corrections to the exponents $\eta$ and $z$ are plotted in the inset. Importantly, there appears to be a discontinuity in the values of both exponents at $\s=7/4$; however, it may very well be the case that higher-order corrections in $\epsilon$-expansion make the jump disappear. While the two-loop correction is rather small at the transition near $\s=7/4$, it gives corrections to both exponents towards their short-range values, and specifically brings the model closer to the `relativistic' point where $z=1$, and, by virtue of Eq.~(\ref{Eq. exact eta and z}), also $\eta=1/4$.
  In the context of long-range interacting classical systems (in the absence of the imaginary-time direction), it was shown that there is no such discontinuity \cite{Sak73}, but this result has been the subject of further scrutiny recently, see Refs.~\cite{Picco12,Rajabpour13,Rajabpour15}. Further analytical and numerical work in the quantum context should also be worthwhile.

  Away from criticality, we note that RG does not modify power laws obtained at the quadratic order.
  This indicates that the exponent in $G_\s(R)\sim 1/R^{1+\s}$, computed for the noncritical model at the quadratic order, does not receive any corrections from RG.

\section{Fermions}\label{Sec: Fermions}
In this section, we study the critical and causal properties of a model of spinless fermions on a chain with long-range hopping and pairing terms. To set up the problem, we first introduce a short-range quadratic lattice model
\begin{equation}\label{Eq. Fermion lattice}
  H=- \sum_i  (c_i^\dagger c_{i+1} +c^\dagger_{i+1}c_i) + J\, (c_i^\dagger c_{i+1}^\dagger +c_{i+1} c_i) -\mu \, c_i^\dagger c_i\,,
\end{equation}
with nearest-neighbor hopping and pairing (with the amplitudes $1$ and $J$, respectively), and a chemical potential $\mu$. The long-wavelength physics of this model is captured by a continuum field theory with the Euclidean action \cite{SachdevBook}
\begin{equation}\label{Eq. action fermions}
  I=\int\!\! d\tau \!\!\int \!\!dx \,\, \Psi^* \frac{\partial \Psi}{\partial \tau} +\frac{1}{2} \left(\Psi^* \frac{\partial \Psi^*}{\partial x} -\Psi \frac{\partial \Psi}{\partial x} \right)+\Delta \, \Psi^* \Psi\,,
\end{equation}
where $\Psi$ is a Grassman field. The spatial coordinate is rescaled to normalize the coefficient of the gradient term, and the constant $\Delta$ is generically related to the parameters in the lattice model; for example, for $J=1$, we have $\Delta=2-\mu$. We shall take the continuum description as our starting point, and are not concerned with the relationship of its parameters to those in the lattice model.
In principle, the Hamiltonian in Eq.\,(\ref{Eq. Fermion lattice}) may also include interaction terms such as $c_i^\dagger c_{i}c^\dagger_{i+1}c_{i+1}$. Mapping to the continuum, such an operator induces an interaction term that must involve spatial gradients (as a result of the Pauli exclusion principle, $\Psi^2=0$). From the scaling dimension of the fermionic field, $[\Psi]=1/2$, one can easily check that all allowable interaction terms are irrelevant in the sense of RG \cite{SachdevBook}, and thus Eq.~(\ref{Eq. action fermions}) is the exact (in the RG sense) continuum description of the lattice model even with interactions.
We also remark that the fermionic field theory in Eq.~(\ref{Eq. action fermions}) is relativistic in the sense that the dynamic exponent is $z=1$, indicating a linear causal region.
The action can be cast in Fourier space as (integral over momentum and frequency suppressed)
\begin{equation}\label{Eq. kernel fermions}
  \begin{pmatrix}
      \Psi_{\omega, q}^* & \Psi_{-\omega,-q}
    \end{pmatrix}\!
    \begin{pmatrix}
      -i\omega+\Delta  & i q  \\
      -i  q & \hskip -.08in -i\omega -\Delta
    \end{pmatrix}\!
    \begin{pmatrix}
      \Psi_{\omega, q}  \\
      \Psi_{-\omega,-q}^*
    \end{pmatrix},
\end{equation}
where $\omega$ is the {Fourier variable corresponding to} the imaginary time. Setting the determinant of the above matrix to zero, we find the dispersion relation in imaginary frequency $\omega=i\omega(q)$ with
$\omega(q)=\sqrt{\Delta^2+ q^2}$.

We are interested in the consequences of adding long-range terms to a fermionic model
in Eq.~(\ref{Eq. Fermion lattice}).
For example, one can add both long-range hopping and long-range pairing terms,
\begin{align}
\mbox{Hopping}:  &   \qquad  \sum_{i \neq j} \frac{1}{|i-j|^{1+\s}} \, c_i^\dagger c_j\,, \nonumber \\
\mbox{Pairing}: &    \qquad  \sum_{i>j} \frac{1}{|i-j|^{1+\s}} \, c_i^\dagger c_j^\dagger + {\rm h.c.} \nonumber
\end{align}
Again, we shall assume $\s>0$---for $-1<\sigma<0$, our treatment requires special care, and will be extended in future work.
Long-range interactions, beyond the quadratic order in fermionic operators, can also be considered, though they do not affect the possible RG-relevant contributions to the continuum theory. Mapping these expressions to the continuum, the matrix in Eq.~(\ref{Eq. kernel fermions}) becomes
\begin{equation}\label{Eq. Fermions LR}
    \begin{pmatrix}
      -i\omega+\Delta-  q^2 +{\cal H}_\s(q) & i q + {\cal P}_\s(iq)  \\
      -i  q+{\cal P}_\s(-iq)        & \hskip -.15in -i\omega -\Delta+  q^2 -{\cal H}_\s(q)
    \end{pmatrix},
\end{equation}
in which ${\cal H}_\s(q)\sim |q|^\sigma$ and ${\cal P}_\s(iq)\sim e^{\pm i\pi\s/2}{|q|}^\sigma$ (with $\pm$ corresponding to $q\gtrless 0$) denote low-momentum expansion of the long-range hopping and pairing terms, respectively.
In the case of long-range hopping, ${\cal P}_\s(iq)\sim \sum_{r=1}^\infty e^{i q r} /r^{1+\s}$, which also diverges for odd-integer values of $\s$. In this case, we shall restrict ourselves to non-integer $\s>0$. In the above equation,
we have also included a quadratic term in $q$ (with a normalized coefficient) on the diagonal, even though it is irrelevant compared to the linear-momentum term; the former will be important in some cases at criticality ($\Delta=0$). The dispersion relation is now given by $\omega_\s(q)=\sqrt{[\Delta-q^2+{\cal H}_\s(q)]^2+|iq+{\cal P}_\s(iq)|^2}$.

Comparing the exponents of the momentum-dependent terms in Eq.~(\ref{Eq. Fermions LR}), we find that long-range terms ($\sim |q|^\s$) are either relevant or irrelevant compared to the linear-momentum term ($\sim q$) depending on whether $\sigma<1$ or $\sigma>1$, respectively. Therefore, the dynamic critical exponent is $z=\s$ when $0<\sigma<1$, and $z=1$ for $\sigma>1$. Furthermore, the scaling dimension of the fermionic field is fixed by the linear time derivative [the first term in Eq.~(\ref{Eq. action fermions})] and  remains the same, $[\Psi]=1/2$, independent of $\s$. Hence, by the same arguments that we applied to the local model, all possible long-range non-linear terms are irrelevant.  The scaling dimension of the Fermi field is discussed in more detail in Sec.~\ref{Sec. Fermions sigma <1}.

Next we define the two-point functions of the Fermi field in the continuum in simple analogy to their lattice definitions, Eqs. (\ref{Eq. G 11 fermion}) and (\ref{Eq. D 11 fermion}). The correlation function is given by
\begin{equation}\nonumber
  i\left[{\mathbf G}_{\s}(t,x)\right]_{\alpha\beta} \propto
   \left\langle {\cal T}\hat \Psi_\alpha(t,x)  \hat \Psi_\beta^\dagger(0,0) \right\rangle,
\end{equation}
where $\hat \Psi$ and $\hat\Psi^\dagger$ are the operator-valued Fermi fields in the Heisenberg picture.
We recall the definitions $\left(\hat\Psi_1 \,\, \hat\Psi_2\right )=\left(\hat\Psi \,\, \hat\Psi^\dagger\right)$ and $\alpha, \beta\in \{1,2\}$. Similarly, the response function is defined as
\begin{equation}\nonumber
   i\left[{\mathbf D}_{\s}(t,x)\right]_{\alpha\beta} \propto \Theta(t)\left\langle \,[\hat\Psi_\alpha(t,x), \hat\Psi^\dagger_\beta(0,0)]_+ \!\right\rangle.
\end{equation}
In the above equations, we have only given the proportionality relations, as we are only interested in scaling relations and not the (nonuniversal) coefficients.

With the fermionic action being quadratic, the two-point functions can be extracted from the dispersion relation. The correlation function at equal times, ${\mathbf G}_\s(R)\equiv  {\mathbf G}_\s(t\to 0,R)$ (with $R=|x|$ the distance between the two points), can be written as \footnote{A careful evaluation of the limit $t \to 0^+$ shows that the first derivative in time only gives $\delta(R)$, which does not contribute to the asymptotic long-distance behavior.}
\begin{equation}\label{Eq. GF}
   {\mathbf G}_{\s}(R)\propto
   \begin{pmatrix}
      \Delta+\partial_R^2  & \partial_R \\
      -\partial_R &  -\Delta-\partial_R^2
   \end{pmatrix}
   \tilde G_\s(R) + \tilde {\mathbf G}_\s(R)\,.
 \end{equation}
 Here $\tilde G_\s(R)$ is the correlation function defined in Eq.~(\ref{Eq. corr fn}), but with $\omega_\s(q)$ now being the dispersion of the fermionic model, and the dispersion $\omega_\s(q)$ of the fermionic model,} and
 \begin{equation}\label{Eq. mathbf G}
   \tilde {\mathbf G}_\s(R)\propto
   \frac{1}{2}\int_{-\infty}^\infty \frac{dq}{\omega_\s(q)} \, e^{i q R}
   \begin{pmatrix}
      {\cal H}_\s(q)  & {\cal P}_\s(iq) \\
      {\cal P}_\s(-iq) & -{\cal H}_\s(q)
   \end{pmatrix}.
 \end{equation}
As before, we have given the proportionality relations as we are interested only in scaling relations and not the precise coefficients [the factor of 1/2 in Eq.~(\ref{Eq. mathbf G}) is relative to the first term on the right-hand side of Eq.~(\ref{Eq. GF})]. In the above equations, the diagonal matrix elements give the normal correlation function ($\langle \Psi \Psi^\dagger\rangle$ and its complex conjugate), while the off-diagonal matrix elements define the anomalous correlation function ($\langle \Psi \Psi\rangle$ and its complex conjugate). Similarly, the response function can be written as
  \begin{equation}\label{Eq. DF}
   {\mathbf D}_{\s}(t,R)\propto
   \begin{pmatrix}
      i\partial_t+\Delta +\partial_R^2 & \partial_R \\
      \hskip -.1in -\partial_R & \hskip -.2in i\partial_t -\Delta-\partial_R^2
   \end{pmatrix}
   \tilde D_\s(t,R) + \tilde {\mathbf D}_\s(t,R)\,,
 \end{equation}
 where $\tilde D_\s(t,R)$
 is the response function defined in Eq.~(\ref{Eq. res fn}), but with $\omega_\s(q)$ now being the dispersion of the fermionic model, and to the dispersion relation $\omega_\s(q)$ of the fermionic model, and
 \begin{align}
   &\tilde {\mathbf D}_\s(t,R)\propto  \nonumber \\
   &\frac{1}{2}\int_{-\infty}^\infty \frac{dq}{\omega_\s(q)} \sin(\omega_\s(q) t)e^{i q R}
   \begin{pmatrix}
      {\cal H}_\s(q)  & {\cal P}_\s(iq) \\
      {\cal P}_\s(-iq) & -{\cal H}_\s(q)
   \end{pmatrix}.
 \end{align}

For the sake of comparison, we first quote the two-point functions in the absence of long-range terms at or away from criticality. Within the short-range model, these functions are directly related to their counterparts in the case of the scalar field
via the first terms in Eqs.~(\ref{Eq. GF}) and (\ref{Eq. DF}).
The correlation function at criticality, $\Delta=0$, is
\begin{equation}\label{Eq. Correlation SR fermions}
  {\mathbf G}_{\rm SR} (R)\sim
  \begin{pmatrix}
      \partial_R^2  & \partial_R \\
      -\partial_R & -\partial_R^2
   \end{pmatrix} G_{\rm SR}(R)\sim
   \begin{pmatrix}
      -1/R^2  & 1/R \\
      -1/R & 1/R^2
   \end{pmatrix},
\end{equation}
where, in the last step, we used $\tilde G_{\rm SR}(R)\equiv G_{\rm SR}(R)\sim \log R$. Away from criticality, the correlation function falls off exponentially beyond the correlation length.
Similarly, the response function at criticality is given by
\begin{equation}
  {\mathbf D}_{\rm SR}(t, R)\sim
  \begin{pmatrix}
      i\partial_t & \partial_R \\
      -\partial_R & i\partial_t
   \end{pmatrix} \Theta(t-R)\sim \delta(t-R)\,,
\end{equation}
indicating that the response to an infinitesimal perturbation travels at the speed of `light'. Note that we have dropped the second spatial derivatives; this and higher-order derivatives may be included, but will only slightly smear out the delta function. Away from the critical point, the response function remains zero outside the light cone, but inside the light cone it has a complicated form given by the first term in Eq.~(\ref{Eq. DF}), where ${\tilde D}_{\rm SR}(t,R)$ should be substituted from Eq.~(\ref{Eq. D short-range}) at $\varrho>0$ with $\sqrt{\varrho}\to \Delta$. In what follows, we undertake a detailed study of the fermionic field theory with long-range hopping and pairing, which is rather rich despite the absence of interactions. We will explore the linear/non-linear causal behavior, expressing it in terms of explicit scaling relations whenever possible.
In all the cases studied below, we shall restrict ourselves to the vicinity of the critical point, where the correlation length is large compared to the lattice spacing, hence the validity of a continuum description.

\subsubsection{$0<\sigma<1$ at criticality}\label{Sec. Fermions sigma <1}
For the critical model, we have $\Delta=0$. We study the two-point functions for the long-range hopping and pairing terms separately.

{\emph{Hopping}.---}In this case, the dispersion relation reads approximately $\omega_\s(q) =\sqrt{|q|^{2\s}+q^2}\approx |q|^\s(1+\frac{1}{2}|q|^{2-2\s})$. Using Eq.~(\ref{Eq. GF}), one can show that the correlation function behaves asymptotically as
\begin{equation}\label{Eq. Correlation fn hopping}
 {\mathbf G}_{\s}(R)\sim
     \begin{pmatrix}
      1/R^{3-2\s}  & 1/R^{2-\s} \\
      -1/R^{2-\s} & -1/R^{3-2\s}
   \end{pmatrix}.
\end{equation}
Note that we have not kept track of the precise coefficients.
At long distances, the dominant power-law is given by $ {1}/{R^{2-\s}}$.
The corresponding exponent, $2-\sigma$, is not consistent with the scaling dimension of the Fermi field, $[\Psi]=1/2$; however, we have dropped a leading-order delta function $\delta(R)$ which has the correct scaling dimension, but nevertheless is purely local. We remark that, even if one reads off the scaling dimension of the Fermi field from the exponent in Eq.~(\ref{Eq. Correlation fn hopping}), interaction terms still would be irrelevant, and our field theory up to the quadratic order in the fermionic field is exact in the RG sense.

To find the response function, we shall focus on the region well outside the linear light cone $t=R$, where one can truncate the dispersion relation at the leading order as $\omega_\s(q)\approx |q|^\s$. Using Eq.~(\ref{Eq. DF}), one finds
\begin{align}\label{Eq. tilde g}\nonumber
   {\mathbf D}_{\s}(t,R) &\sim
   \begin{pmatrix}
      i\partial_t & \partial_R \\
       -\partial_R & i\partial_t
   \end{pmatrix}
   \left[\frac{1}{R^{1-\s}} \, { h}_{\s}\!\!\left(\frac{t}{R^\s}\right)\right]\nonumber \\
   &+ \begin{pmatrix}
      1 & 0 \\
       0 & -1
   \end{pmatrix}\frac{1}{R} \, {\tilde h}_\s\!\!\left(\frac{t}{R^\s}\right).     \nonumber
\end{align}
The first term in this equation is obtained by acting with linear derivatives on the response function computed in Eq.~(\ref{Eq. fn g}) with ${h}_\s(s)\equiv g_{2\sigma}(s)$ [the function $g_\s$ is defined in Eq. (\ref{Eq. fn g})], while the second term is due to $\tilde {\mathbf D}_\s$ in Eq.~(\ref{Eq. DF}), with $\tilde h_\s$ defined as
\begin{equation}
  \tilde h_\s(s)=\int_0^\infty dq \sin\left(q^\s s\right)\cos q\,.
\end{equation}
The derivatives in the above expression for ${\mathbf D}_{\s}$ can be organized to cast it in terms of two separate scaling functions
\begin{align}\label{Eq. DF for hopping}
  {\mathbf D}_{\s}(t,R) \sim
  \begin{pmatrix}
       \frac{1}{R} \, h_{11} \!\!\left(\frac{t}{R^\s}\right)& \frac{1}{R^{2-\s}} \, h_{12} \!\!\left(\frac{t}{R^\s}\right) \\ \\
       -\frac{1}{R^{2-\s}}  \,h_{12} \!\!\left(\frac{t}{R^\s}\right)  & -\frac{1}{R} \,h_{11}^{*} \!\!\left(\frac{t}{R^\s}\right)
   \end{pmatrix},
\end{align}
where
\begin{align}
  h_{11}(s)&=i\,h'_\s \!\!\left(s \right)+\tilde h_\s \!\left(s\right)\,, \nonumber \\
h_{12}(s)&=-(1-\s)h_\s \!\left(s\right)-\s s \, h_\s'\!\!\left(s\right)\,,
 \end{align}
and the prime on $h_\s$ indicates the first derivative.
Thus all the terms in the response function feature non-linear causal regions with the relative scaling of time and space coordinates satisfying
\begin{equation}\label{Eq. z fermions}
  t\sim R^\sigma,
\end{equation}
consistent with the dynamic critical exponent $z=\s$.
The dominant contribution at long times and distances is given by the diagonal terms of Eq.~(\ref{Eq. DF for hopping}) and is of the form
\begin{equation}\label{Eq. DF scaling form}
  {\mathbf D}_{\s}(t,R) \sim \frac{1}{R} \, {\mathbf H}_\s\!\!\left(\frac{t}{R^\s}\right),
\end{equation}
with
\begin{equation}
  {\mathbf H}_\s(s)=\begin{pmatrix}
        h_{11}(s)& 0\\
       0 & h_{22}(s)
   \end{pmatrix}.
\end{equation}
Finally, at long distances and at a fixed time, one can see that the dominant term is $(1/R)\,\tilde h\left(t/R^\s\right) \sim t/R^{1+\s}$ [since $\tilde h(s)\sim s$ for small $s$], again consistent with the Hastings-Koma bound \cite{hastings05} for the lattice model of fermions with long-range hopping.

{\emph{Pairing}.---}In this case, the dispersion reads $\omega_\s(q) =\sqrt{|q|^{2\s}+|q|^{1+\s}}\approx |q|^\s(1+\frac{1}{2}|q|^{1-\s})$. Using Eq.~(\ref{Eq. GF}), the correlation function becomes
\begin{equation}\label{Eq. Correlation fn pairing}
 {\mathbf G}_{\s}(R)\sim
     \begin{pmatrix}
      1/R^{3-\s}  & 1/R^{2-\s} \\
      -1/R^{2-\s} & -1/R^{3-\s}
   \end{pmatrix}.
\end{equation}
At long distances, the dominant power-law is given by $ {1}/{R^{2-\s}}$. We refer to the paragraph following Eq.~(\ref{Eq. Correlation fn hopping}) regarding the associated exponent in comparison with the scaling dimension of the Fermi field.

Similar to the hopping case, in order to find the response function in the region well outside the linear light cone $t=R$, we can approximate the dispersion relation as $\omega_\s(q)\approx |q|^\s$. One can go through the same argument as above to obtain the asymptotic (at large distances and times) response function as
  \begin{align}\label{Eq. DF for pairing}
  {\mathbf D}_{\s}(t,R) \sim
  \frac{1}{R}\begin{pmatrix}
        k_{11} \!\!\left(\frac{t}{R^\s}\right)& k_{12} \!\!\left(\frac{t}{R^\s}\right) \\ \\
       -k_{12} \!\!\left(\frac{t}{R^\s}\right)  & k_{11} \!\!\left(\frac{t}{R^\s}\right)
   \end{pmatrix},
  \end{align}
  where
  \begin{align}
  k_{11}(s)&=i\,h'_\s \!\!\left(s \right), \nonumber \\
k_{12}(s)&= \tilde k_\s(s),
 \end{align}
and ${\tilde k}_\s$ is defined as
\begin{equation}
  \tilde k_\s(s)=\int_{0}^\infty dq \sin\left(q^\s s\right)\cos(q +\pi\s/2)\,.
\end{equation}
Again, we find a nonlinear causal behavior described by Eqs.~(\ref{Eq. z fermions}) and (\ref{Eq. DF scaling form}) with the identification
\begin{equation}
  {\mathbf H}_\s(s)=
  \begin{pmatrix}
        k_{11} \!\!\left(s\right)& k_{12} \!\!\left(s\right) \\
       -k_{12} \!\!\left(s\right)  & k_{11} \!\!\left(s\right)
   \end{pmatrix}.
\end{equation}
Furthermore, the scaling with $R$ at fixed $t$ is consistent with the Hastings-Koma bound applied directly to the lattice model of fermions with long-range pairing (as can be seen by noting that, similar to $\tilde h_\s(s)$, we have $\tilde k_\s (s)\sim s$ for small $s$).

\subsubsection{$\sigma>1$ at criticality}
For $\s>1$, the dispersion relation can be approximated at small $|q|$ as $\omega_\s(q) \approx |q|$.
The correlation function is then given by
\begin{equation}\nonumber
  {\mathbf G}_{\s}(R) \sim \int \frac{dq}{q}
   \begin{pmatrix}
      q^2 + {\cal H}_\s(q)  & iq+{\cal P}_\s(iq) \\
      -iq + {\cal P}_\s(-iq) & -q^2-{\cal H}_\s(q)
   \end{pmatrix}e^{iqR}.
\end{equation}
In the case of long-range hopping, we find
\begin{equation}
  {\mathbf G}_{\s}(R) \sim
  \begin{pmatrix}
      1/R^\alpha  & 1/R \\
      -1/R & -1/R^\alpha
   \end{pmatrix},
\end{equation}
with $\alpha=\min(2,\sigma)$. The dominant power-law at long distances is given by $1/R$.
In the case of long-range pairing, we find the same asymptotic dependence on $R$ as in the short-range model, Eq.~(\ref{Eq. Correlation SR fermions}).

The response function can be computed with the aid of Eq.~(\ref{Eq. DF}), the first term of which simply consists of derivatives acting on $\tilde D_\s(t,R)$ from the scalar case. The dispersion relation to the subleading order takes the form $\omega_\s(q)\approx |q|+B_\alpha|q|^{\alpha-1}$ with $\alpha =\min(2\s, 2+\s)$ and some constant $B_\alpha$. Therefore, the function $\tilde D_\s(t,R)$ can be taken from Eq.~(\ref{Eq. Scaling eq}) with the identification $\sigma \to \alpha =\min(2\s, 2+\s)$, that is $\tilde D_\s(t,R)=D_\alpha(t,R)$. Notice that $\alpha>2$ for $\s>1$, hence, $\tilde D(t,R)$ in the first term of Eq.~(\ref{Eq. DF}) exhibits a linear causal behavior when $\s>1$; derivatives acting on this term should not change the causal behavior. Also, the second term in Eq.~(\ref{Eq. DF}), $\tilde {\mathbf D}_\s(t,R)$, can be computed asymptotically at long times as (assuming $\omega_\s(q) \approx |q|$ and $R>t$)
\begin{align}
  \tilde {\mathbf D}_\s(t,R)&=
  \int dq \, q^{\s-1} \sin(qt)\cos(q R)\nonumber \\
  &\sim \frac{1}{(R-t)^\s}- \frac{1}{(R+t)^\s}\,.\nonumber
\end{align}
Near the light cone $t=R$, we have $\tilde {\mathbf D}_\s(t,R) \sim 1/(R-t)^\s$ at long times, indicating a linear causal behavior. Therefore, the full response function ${\mathbf D}_{\s}$ indeed gives rise to a linear light cone.
Furthermore, at long distances and fixed time, $\tilde{\mathbf D}_\s(t,R)\sim t/R^{1+\s}$ is again consistent with the Hastings-Koma bound \cite{hastings05} for the corresponding lattice model.

For long-range pairing, the dispersion relation to the subleading order in $|q|$ is given by $\omega_\s(q)=|q| +B'_\alpha|q|^{\alpha-1}$ with $\alpha=\min(2\s,1+\s)>2$ and some constant $B'_\alpha$; once again $\alpha>2$ for $\s>1$. One can argue, in a similar manner to the case of long-range hopping above, that the causal region is bounded by a linear light cone.

\subsubsection{All $\sigma>0$ away from criticality}
For long-range hopping, and setting $\Delta=1$ for convenience, the dispersion relation is given by $\omega_\s(q)=\sqrt{1+q^2+|q|^\s}$, which yields, via Eq.~(\ref{Eq. GF}),
\begin{equation}\label{Eq. corr fn-hopping}
  {\mathbf G}_{\s}(R)\sim
  \begin{pmatrix}
      1/R^{1+\s}  & 1/R^{2+\s} \\
      -1/R^{2+\s} & -1/R^{1+\s}
   \end{pmatrix}.
\end{equation}
At long distances, the dominant power law is given by the diagonal contributions $\sim{1}/{R^{1+\s}}$.

For long-range pairing, the dispersion relation is given by $\omega_\s(q)=\sqrt{1+q^2+|q|^\alpha}$ with $\alpha=\min(2\s,1+\s)$; we find
\begin{equation}\label{Eq. corr fn-pairing}
{\mathbf G}_{\s}(R)\sim \begin{pmatrix}
      1/R^{1+\alpha}  & 1/R^{1+\s} \\
      -1/R^{1+\s} & -1/R^{1+\alpha}
   \end{pmatrix},
\end{equation}
in agreement with Ref.~\cite{Pupillo14}.  At long distances, the dominant power law comes from the off-diagonal contributions $\sim{1}/{R^{1+\s}}$.
We note that, in both cases above, the correlation functions are described by the same asymptotic power-law as that of the long-range couplings in the lattice Hamiltonian. While this appears to be a fairly generic feature of long-range interacting models, there are exceptions \cite{Gong15}.

The response function is given by Eq.~(\ref{Eq. DF}) and can be cast as
\begin{equation}\label{Eq. noncritical fermions hopping}
  {\mathbf D}_{\s}(t,R) \sim  \pm \tilde D_\s(t,R) + \mbox{higher derivatives}\,,
\end{equation}
where those terms that include higher derivatives or higher factors of momenta in the integrand ($\tilde{\mathbf D}_\s$) are neglected.
Therefore, information about causality in the non-critical fermionic system is captured by $\tilde D_\s(t,R)$, and is essentially identical to that of a scalar-field model with the same dispersion.
Consequently, the conclusions of Sec.~\ref{Sec. All sigma noncritical} should be immediately applicable here.
Specifically, for long-range hopping, $\omega_\s(q)=\sqrt{1+q^2+|q|^\s}$, and we find a non-linear light-cone for $\sigma<1$.
On the other hand, for long-range pairing, $\omega_\s(q)=\sqrt{1+q^2+|q|^\alpha}$ with $\alpha=\min(2\s, 1+\s)$. The nonlinear light cone then arises for $\alpha<1$, or equivalently $\s<1/2$.

\section{Conclusion and Outlook}
In this work, we have studied the critical and near-critical properties of the long-range interacting TFIM and of a fermionic model with long-range hopping and pairing. We have identified the critical exponents characterizing the dynamic behavior as well as the two-point correlation functions of the system. For the TFIM, a nontrivial calculation gives the value of these exponents to the two-loop order.  Field-theoretical arguments indicate that the fermionic model is exact (in the RG sense) at the quadratic level, from which we are able to calculate various critical exponents.  At the critical point
, we have argued that the causal behavior is identified by the dynamic critical exponent, and the response functions find a general scaling form, explicit and universal properties of which have been obtained. For both critical and noncritical models, we have derived the regimes of linear/non-linear light cones, mostly with the aid of the dynamic critical exponent.

Our definition of the causal behavior and causal region is inspired by field theory; specifically, the linear causal region at criticality is identified with $z=1$, indicating relativistic dynamics. We have argued that values of $z<1$ correspond to a sublinear causal region.  It is shown that the causal region defined by the dynamic critical exponent indeed characterizes the local maxima of the response function, suggesting a close connection to the speed of information propagation. We have also shown that the response function can generally be expressed in terms of scaling functions up to multiplicative decaying power laws.  Recalling that, in a local quench, the signal typically decays with distance, we find our definition of the causal region based on the scaling function quite natural.  However, we stress that this definition need not produce the same light cone that would be obtained by setting the tightest possible Lieb-Robinson-type bound equal to a constant, a subject that merits further investigation.  If the two agree at criticality, where our calculations predict the strongest deviations from a linear light cone, we conjecture that the TFIM and the Fermion Model exhibit a linear causal behavior for $1+\s\gtrapprox 3$ and $1+\s\ge2$, respectively. {This conjecture is particularly valuable in light of the fact that even the tightest available Lieb-Robinson-type bound for long-range interacting systems cannot exclude the existence of a nonlinear light cone at any finite $\sigma$, no matter how large \cite{Foss-Feig15}.
It would also be worthwhile to explore possible connections between this conjecture and the proof that the light cone cannot be logarithmic, and is at worst algebraic, for $1 + \sigma > 2$ \cite{Foss-Feig15}.

The powerful tools of field theory used in this paper can be employed to study a range of interesting, and experimentally relevant, long-range interacting systems \cite{baranov12,pohl11,gopalakrishnan11,douglas15,blatt12},
such as a huge variety of spin-1/2 \cite{micheli06,gorshkov11b,gorshkov11c,manmana13}, spin-1 \cite{brennen07,cohen15b,Gong15}, and higher-spin \cite{barnett06,gorshkov13}  models,
 generalized Hubbard \cite{barnett06,wall13b,dutta15} and $t$-$J$ models \cite{gorshkov11b,gorshkov11c},
  and spin-boson problems \cite{junemann13}, among many others, in one or more spatial dimensions. In general, these models exhibit new universal behavior not captured by standard long-range interacting classical models, since the quantum-to-classical mapping generates classical models with long-range interactions in all spatial directions except the one corresponding to the imaginary time dimension of the quantum model \cite{Amit01}.
Another interesting, but more complicated, direction is to study the consequences of a global quench as opposed to a local perturbation.
However, in this case, it is not at all obvious that field-theorertic considerations will be valid given the extensive energy imparted by the quench (though see Ref.~\cite{PhysRevA.90.063622} for related studies of weak global quenches).
Finally, this entire line of speculation is intimately related to the growth and propagation of entanglement in systems with long-range interactions \cite{hastings05,hauke13,Schachenmayer13,Rajabpour13-2,Nezhadhaghighi14}, a subject for which our understanding is far from complete.

\begin{acknowledgments}
We thank M.\ A.\ Rajabpour, L.\ Lepori, G.\ Pupillo, and D.\ Vodola for discussions. This work was supported by the NSF PIF, AFOSR, ARO, ARL, NSF PFC at the JQI, and AFOSR MURI. M. F.-F. thanks the NRC for support.
\end{acknowledgments}

\appendix
\section{Two-point functions in quadratic scalar and fermionic field theories}\label{App. 2pt functions}
In this appendix, we derive the analytical expressions for the two-point functions in a quadratic field theory in terms of the dispersion relation $\omega(q)$.

 \subsection{Scalar field}\label{Sec. Saclar field}
 For a quadratic scalar field theory, the imaginary-time-ordered correlation function in Fourier space is given by
\begin{equation}\label{Eq. cal G}
    {\cal G}(\omega, q)=\frac{1}{\omega^2+ \omega(q)^2}\,,
\end{equation}
  with $\omega(q)$ the dispersion relation.

\

  \textbf{Correlation function.---}The time-ordered correlation function in real time, $G(t,R)$, is related to $\cal G$ (in imaginary time and real space) as \cite{WenBook}
 \begin{equation}
   i G(t,R)={\cal G}(it+\sgn(t) 0^+,R)\,.
 \end{equation}
 Thus,
 \begin{align}\nonumber
   iG(t \to 0,R)&={\cal G}(0,R)\propto\int {d\omega dq}\,{\cal G}(\omega,q)e^{i q R},
 \end{align}
 which leads to Eq.~(\ref{Eq. corr fn}) for the correlation function at equal times.

\

  \textbf{Response function.---}The causal response function  $D(t,R)$ is related  to ${\cal G}$ via an analytic continuation in frequency space as $i\omega \to \omega + i0^+$ \cite{WenBook},
   that is
  \begin{equation}
    D(\omega,q)=-{{\cal G}(\omega,q)}{\Big |}_{i\omega\to \omega+i0^+}=\frac{1}{(\omega+i0^+)^2- \omega(q)^2}\,.
  \end{equation}
  This equation, cast in real space and time, leads to Eq.~(\ref{Eq. res fn}).
\
 \subsection{Fermionic field}\label{Sec. Fermionic field}
 For the fermionic field, the imaginary-time-ordered correlation function is obtained by inverting the $2\times 2$ matrix in Eq.~(\ref{Eq. Fermions LR}),
 \begin{align}\label{Eq. Green fn fermion}
   {\cal G}_F(\omega, q)&=
   \begin{pmatrix}
      i\omega+\Delta-q^2  & i q  \\
      -i  q & i\omega -\Delta+q^2
   \end{pmatrix}
   {\cal G}(\omega,q)  \nonumber \\
   &+
   \begin{pmatrix}
      {\cal H}(q)  & {\cal P}(iq) \\
      {\cal P}(-i q) & -{\cal H}(q)
   \end{pmatrix}
   {\cal G}(\omega,q),
 \end{align}
 with ${\cal G}$ from Eq.~(\ref{Eq. cal G}) (with $\omega(q)$ specific to the fermionic theory and the subscript $\s$ dropped). The first line of this equation can be cast in terms of the function ${\cal G}$ and its spatial/tempoal derivatives, while the second line cannot.
 A similar analytical continuation described for the scalar case leads to Eqs.~(\ref{Eq. GF}) and (\ref{Eq. DF}), wherein the tilde functions correspond to the second line of Eq.~(\ref{Eq. Green fn fermion}).

\

  \section{RG for the $\phi^4$ model with long-range interactions in the spatial direction}\label{App. RG}
  In this appendix, we perform the RG calculation including the two-loop diagrams for the $\phi^4$ model {in 1+1 dimensions with long-range interactions along the spatial direction. Although we are interested in $d=1$ spatial dimension, we generalize to a $d$-dimensional space in order to expand around the upper critical dimension $d=d_u$.}
  We first cast the action in imaginary frequency and momentum space as
\begin{widetext}
\begin{align}\label{Eq. Hamiltonian in Fourier}
  H&=\int \frac{d^d\bq}{(2\pi)^d}\int \frac{d\omega}{2\pi}\frac{\varrho+A\omega^2+B_\sigma q^\sigma}{2}\phi(\omega,\bq)\phi(-\omega,-\bq) \nonumber \\
  &+u \int\frac{d^d\bq_1d^d\bq_2d^d\bq_3}{(2\pi)^{3d}}\int\frac{d\omega_1d\omega_2d\omega_3}{(2\pi)^3}\phi(\omega_1,\bq_1)
  \phi(\omega_2,\bq_2)\phi(\omega_3,\bq_3)\phi(-\omega_1-\omega_3-\omega_3,-\bq_1-\bq_2-\bq_3),
\end{align}
\end{widetext}
where $q=|\bq|$; note that we are using a different normalization (a factor of 1/2) for the quadratic terms in $\phi$ for convenience. (In a slight abuse of notation, we use the same coefficients $A$, $B_\s$, and $\varrho$ defined in the action cast in real space and time coordinates.)
In order to compute loop diagrams, we rely upon the momentum-shell RG \cite{Wilson74}. However, similar to Ref.\ \cite{Hohenberg77}, we do not find it necessary to introduce a cutoff in frequencies, and will integrate over $\omega\in (-\infty,\infty)$, which is free of divergences.
Without long-range interactions, time and space coordinates appear on equal footing in the action [$(\partial_\tau \phi)^2+v^2(\nabla \phi)^2$ for some velocity $v$], and long wave-length physics becomes `Lorentz' invariant, which yields $z=1$ exactly. For long-range interactions, on the other hand, there is no such symmetry, and it is natural to treat space and time coordinates differently as prescribed in the main text.

We first compute the one-loop correction (depicted in Fig.~\ref{Fig. 2ptLoops}) to the parameter $\varrho$ in the action;
we find
\begin{align}
   &{4 \choose 2} u \int_{\Lambda/b<|\bq|<\Lambda} \frac{d^d\bq}{(2\pi)^d} \int_{-\infty}^{\infty}\frac{d\omega}{2\pi}\frac{1}{\varrho+A\omega^2+B_\sigma q^\sigma} \nonumber \\
  =&{4 \choose 2}\frac{u}{2\sqrt{A}} \int_{\Lambda/b<|\bq|<\Lambda} \frac{d^d\bq}{(2\pi)^d} \frac{1}{\sqrt{\varrho+B_\sigma q^\sigma}}\,, \nonumber
\end{align}
where we have used the  identity $\int_{-\infty}^\infty\frac{d\omega}{2\pi} \frac{1}{A\omega^2+B}=\frac{1}{2\sqrt{AB}}$. The constant $\Lambda$ is the momentum cutoff in the model, and the integral is over the momentum shell $\Lambda/b<|\bq|<\Lambda$. We specialize to an infinitesimal rescaling $b=1+\delta l$ and expand everything to  first order in $\delta l$. The RG flow of $\varrho$, combined with its scaling dimension in Eq.~(\ref{Eq. Engineered scaling dim}), is then given by
\begin{equation}\label{Eq. RG for r}
  \frac{d\varrho}{dl}=(z+d-2a)\varrho+\frac{6 K_d \Lambda^d u }{\sqrt{A}\sqrt{\varrho+B_\sigma \Lambda^\sigma}} + {\cal O}(u^2),
\end{equation}
where $K_d=S_d/(2\pi)^d$, with $S_d$ the surface area of a sphere in $d$ dimensions.

Next we compute the one-loop RG flow of the interaction vertex.
In computing the corresponding diagram, we drop the dependence on external momenta and frequencies which otherwise produce derivatives in the nonlinear terms, and therefore can be neglected \cite{KardarBook}.
We obtain, for the one-loop diagram,
\begin{align}
  & {4 \choose 2}{4 \choose 2}{u^2} \int_{\Lambda/b<|\bq|<\Lambda} \frac{d^d\bq}{(2\pi)^d} \int_{-\infty}^{\infty}\frac{d\omega}{2\pi} \frac{1}{\left(\varrho+A\omega^2+B_\sigma q^\sigma\right)^2} \nonumber \\
  =&\,\frac{9u^2}{\sqrt{A}}\int_{\Lambda/b<|\bq|<\Lambda} \frac{d^d\bq}{(2\pi)^d} \frac{1}{\left(\varrho+B_\sigma q^\sigma\right)^{3/2}}\nonumber\,,
\end{align}
where we have used the identity $\int_{-\infty}^\infty\frac{d\omega}{2\pi} \frac{1}{(A\omega^2+B)^2}=\frac{1}{4\sqrt{AB^3}}$.
The RG flow for $u$, combined with Eq.~(\ref{Eq. Engineered scaling dim}), reads
\begin{equation}\label{Eq. RG for u}
  \frac{du}{dl}= (z+d-4a)u-\frac{9 K_d \Lambda^d u^2}{\sqrt{A}(\varrho+B_\sigma \Lambda^\sigma)^{3/2}} +{\cal O}(u^3)\,.
\end{equation}
At the one-loop order, neither $A$ nor $B_\sigma$ are renormalized, and thus their RG flow is simply given by the \emph{engineered }scaling dimensions in Eq.~(\ref{Eq. Engineered scaling dim}),
\begin{align}\label{Eq. Engineered scaling dim in RG}
  \frac{dB_\sigma}{dl}&= (z-\sigma+d-2a)B_\sigma +{\cal O} (u^2)\,, \nonumber \\
  \frac{dA}{dl}&=(-z+d-2a)A+{\cal O}(u^2)\,.
\end{align}
To this order, the above set of equations determine $a=(d-\sigma/2)/2$ and $z=\sigma/2$ (identical to the mean-field values).
The RG equations (\ref{Eq. RG for r}) and (\ref{Eq. RG for u}) for $\varrho$ and $u$ then take the form
\begin{align}\label{RG for r and u}
  \frac{d\varrho }{dl}&=\sigma \varrho +\frac{6  K_d \Lambda^d u}{\sqrt{A}\sqrt{\varrho+B_\sigma \Lambda^\sigma}}\,,\nonumber \\
  \frac{du}{dl}&=\epsilon u - \frac{9 K_d \Lambda^d u^2}{\sqrt{A}(\varrho+B_\sigma \Lambda^\sigma)^{3/2}}\,.
\end{align}
The fixed point is obtained by $d\varrho /dl=du/dl=0$, and yields the fixed-point values $u=u_*$ and $\varrho=\varrho_*$ given by
\begin{equation}\label{Eq. u* r*}
  \varrho_*=-\frac{2B_\sigma \Lambda^\sigma}{3\sigma}\,\epsilon,\qquad u_*=\frac{\sqrt{A}B_\sigma^{3/2}}{9K_{d_u}}\, \epsilon,
\end{equation}
where we have kept the dependence on $\epsilon$ only to the first order, thereby replacing other appearances of the dimension $d$ by $d_{u}$.
Now the RG flow near the critical point can be linearized as
\begin{equation}\label{Eq. delta u delta r}
  \frac{d}{dl}
  \begin{pmatrix}
    \delta \varrho \\
    \delta u
  \end{pmatrix}=
  \begin{pmatrix}
    \sigma-\epsilon/3 \quad & \cdots \\
    {\cal O}(\epsilon^2) &   -\epsilon
  \end{pmatrix}
  \begin{pmatrix}
    \delta \varrho \\
    \delta u
  \end{pmatrix},
\end{equation}
where $\delta \varrho$ and $\delta u$ are deviations from their corresponding fixed-point values.
The top element of the second column is not computed as it is not necessary for computing eigenvalues \cite{KardarBook}.
The above eigenvalue equation can be used to compute various critical exponents. For the critical exponent defined via the divergence of the correlation length  near the critical point, $\xi\sim (\delta \varrho)^{-\nu}$, we find
\begin{equation}
  \frac{1}{\nu}=\sigma-\frac{\epsilon}{3}+{\cal O}(\epsilon^2)\,,
\end{equation}
the critical exponent characterizing the divergence of susceptibility, $\chi \sim (\delta \varrho)^{-\gamma}$, is given by
\begin{equation}
  \gamma=1+\frac{\epsilon}{3\sigma}+{\cal O}(\epsilon^2)\,,
\end{equation}
and, the critical exponent $\eta$ characterizing the anomalous decay of the correlation function, $G(R)\sim 1/R^{d-1+\eta}$, becomes
\begin{equation}\label{Eq. eta}
  \eta=1-\sigma/2 + {\cal O}(\epsilon^2),
\end{equation}
to the first order in $\epsilon$-expansion.
Together with the dynamic exponent $z$, the above exponents satisfy the general identity
\begin{equation}\label{Eq. Hyperscaling rel}
  \gamma= \nu (z-\eta+1)\,.
\end{equation}

Next we go to the two-loop order to find corrections to the exponents $z$ and $\eta$.
We first note that knowledge of the fixed-point values of $\varrho$ and $u$ to  first order in $\epsilon$ suffices to find the above exponents at the $\epsilon^2$ order. To this end, we need to compute the two-loop diagram in Fig.~\ref{Fig. 2ptLoops}; we find
\begin{widetext}
\begin{align}\label{Eq. Two-loop diagram}
  3{4 \choose 1}{4 \choose 1}u^2 &
  \int_{\Lambda/b<|\bq|,|\bp|,|\bw|<\Lambda} \frac{d^d\bq \, d^d\bp}{(2\pi)^{2d}} \int_{-\infty}^{\infty}\frac{d\omega' d\omega''}{(2\pi)^2}\nonumber \\
  & \times \frac{1}{\varrho+A{\omega'}^2+B_\sigma q^\sigma}\times\frac{1}{\varrho+A{\omega''}^2+B_\sigma p^\sigma}\times\frac{1}{\varrho+A(\omega'+\omega''-\omega)^2+B_\sigma |\bq+\bp-\bk|^\sigma}\,,
\end{align}
where $\bw=\bq+\bp-\bk$, and $\bk$ and $\omega$ are the external momentum and frequency, respectively.
This expression possibly gives corrections to the frequency- and momentum-dependent terms, i.e. the coefficients $A$ and $B_\sigma$, in the quadratic part of the action (\ref{Eq. Hamiltonian in Fourier}). To find such corrections, we must expand the above integral for small frequencies and momenta keeping in mind that $|\omega|\ll |\omega'|,  |\omega''|$ and $|\bk|\ll |\bq|, |\bp|$. A Taylor expansion in the former will produce only analytical terms, and thus terms such as $q^\sigma$ cannot be generated in this procedure. This implies that $B_\sigma$ is not renormalized, and leads to the non-renormalization condition
\begin{equation}\label{Eq. non-ren condition}
  z-\sigma+d-2 a=0\,.
\end{equation}
This is an exact equation, and is not affected by higher orders of perturbation theory.
Combined with $a=(d-1+\eta)/2$, this equation yields the exact relation between the critical exponents $\eta$ and $z$ in Eq.~(\ref{Eq. exact eta and z}).
In  light of this identity, we find that Eq.~(\ref{Eq. Hyperscaling rel}) takes a particularly simple form,
\begin{equation}
  \gamma=\sigma \nu\,.
\end{equation}

Next, we consider the renormalization of the $\omega^2$ term in the action. We should expand the expression (\ref{Eq. Two-loop diagram}) for the two-loop diagram to second order in $\omega$,
\begin{equation}\label{Eq. Two-loop omega^2}
  {\cal O}(u^2 \omega^0) -\frac{48u^2}{4B_\sigma^3}\,\omega^2 \int_{1/b<|\bq|,|\bp|,|\bw|<1} \frac{d^{d_u}\bq \, d^{d_u}\bp}{(2\pi)^{2{d_u}}} \frac{1}{q^{\sigma/2}p^{\sigma/2}w^{\sigma/2}}\frac{1}{\left(q^{\sigma/2}+p^{\sigma/2}+w^{\sigma/2}\right)^{3}}\,,
\end{equation}
where we have assumed that $\varrho$ is small, justified by the fact that $\varrho \sim \epsilon$ at the fixed point.
The first term, being constant in $\omega$, renormalizes $\varrho$ to  order ${\cal O}(\epsilon^2)$, but is inconsequential in determining the renormalization of $A$ at this order. With the knowledge that $u_*\sim \epsilon$, we have also replaced $d$ by the upper critical dimension $d_u=3\sigma/2$ in order to keep only the leading order in $\epsilon$. In computing the second term in Eq.~(\ref{Eq. Two-loop omega^2}), we have used the identity
\begin{align*}
&\int_{-\infty}^\infty\frac{d\omega' d\omega''}{(2\pi)^2} \frac{1}{A{\omega'}^2+a^2}\times \frac{1}{A{\omega''}^2+b^2}\times \frac{1}{A(\omega'+\omega''-\omega)^2+c^2}\nonumber \\
&=\frac{a+b+c}{4A a b c\left[(a+b+c)^2+A\omega^2\right]} ={\rm const} -\frac{\omega^2}{4abc (a+b+c)^3}+{\cal O}(\omega^4).
\end{align*}
Also, in the momentum integral, we have rescaled all momenta by $\Lambda$; powers of $\Lambda$ from the integral measure and the integrand cancel each other out. It is worth pointing out that, unlike the one-loop integral, one cannot consider an infinitesimal rescaling. In fact, it is convenient to consider a finite $b$, and reiterate the renormalization group transformation $n$ times.
To evaluate the integral in Eq.\ (\ref{Eq. Two-loop omega^2}), we note that a rigid rotation of all three momenta $\bq$, $\bp$, and $\bw$ does not change the integral. We thus fix, say, $\bq=q \hat {\bz}$ in the direction $\hat \bz$, and cast the integral over $\bq$ as
\(
\int \frac{d^{d_u}\bq}{(2\pi)^d}=K_{d_u} \int dq\, q^{d_u-1}
\). The integral over $\bp$ can be then cast in polar coordinates, in $d_u$ dimensions, as
\(
\int \frac{d^{d_u}\bp}{(2\pi)^{d_u}}=\frac{K_{d_u-1}}{2\pi} \int_0^\pi d\theta (\sin\theta)^{d_u-2} \int dp\, p^{d_u-1}
\). In these coordinates, the integration limits are given by $1/b<q,p<1$, and
\[
\frac{1/b^2-q^2-p^2}{2qp}< \cos\theta<\frac{1-q^2-p^2}{2qp}\,.
\]
A change of variable $p\equiv q y$ allows us to write the momentum integral in Eq.~(\ref{Eq. Two-loop omega^2}) as (with $d_u=3\sigma/2$)
\begin{equation}\label{Eq. I}
  \frac{K_{3\sigma/2} K_{3\sigma/2-1}}{2\pi}\int_{1/b}^1 \frac{dq}{q}\int dy\int d\theta \,\frac{y^{\sigma-1} (\sin \theta)^{3\sigma/2-2}}{(1+y^2+2y\cos \theta)^{\sigma/4}\left[1+y^{\sigma/2}+(1+y^2+2y\cos\theta)^{\sigma/4}\right]^{3}}\,,
\end{equation}
where the limits of integration take the form $1/(bq) <y<1/q$
and a more complicated expression for $\cos\theta$. However, since we are only interested in the leading logarithmic dependence, $\log b$, we can simply replace the limits by $0<y<\infty$ and $0<\theta<\pi$ without encountering any divergences. For example, for $y \to 0$, the integrand goes as $y^{\sigma-1}$, while, for $y\to \infty$, it goes as $1/y^{\sigma+1}$, thereby being perfectly convergent in both cases for $\sigma>0$. This simplification allows one to directly compute the integral over $q$ to obtain a factor of $\log b$. The remaining integral over $y$ and $\theta$ then directly contributes to the two-loop diagram as
\begin{equation}\label{Eq. log b}
  {\rm const}- \omega^2\, (\log b) \, \varsigma(\sigma)\epsilon^2\,,
\end{equation}
where
\begin{equation}\label{Eq. s(sigma)}
  \varsigma(\sigma)=\frac{4 \Gamma(3\sigma/2)}{27\sqrt{\pi}\Gamma(3\sigma/4-1/2)}\int_0^{\infty}dy\int_0^\pi d\theta \,\frac{y^{\sigma-1} (\sin \theta)^{3\sigma/2-2}}{(1+y^2+2y\cos \theta)^{\sigma/4}\left[1+y^{\sigma/2}+(1+y^2+2y\cos\theta)^{\sigma/4}\right]^{3}}\,.
\end{equation}
Note that in computing this expression we have used the fixed-point value of $u_*$ given by Eq.~(\ref{Eq. u* r*}). We have also verified numerically that the exact evaluation of the integral in Eq.~(\ref{Eq. I}) almost exactly reproduces the $\log b$ term in Eq.~(\ref{Eq. log b}).
The recursion relation for $A_n$ is then given by
\begin{equation}
  A_{n+1}= A_n \, b^{\sigma-2z} \left[1+2 (\log b) \, \varsigma(\sigma)\epsilon^2\right]\approx A_n b^{\sigma-2z+2 \varsigma (\sigma)\epsilon^2 },
\end{equation}
where the exponent of $b$ in the first equality is obtained from the scaling dimension in Eq.~(\ref{Eq. Engineered scaling dim}) [or, equivalently, Eq.~(\ref{Eq. Engineered scaling dim in RG})] combined with the non-renormalization condition, Eq.~(\ref{Eq. non-ren condition}), while the expression in the bracket gives the two-loop correction. In the last equality, we have exponentiated the $\epsilon$-dependent term justified by our perturbative treatment in $\epsilon$. The above recursion relation goes to a finite fixed point $A_*$ if the exponent $z=\sigma/2+\Delta z$ is given by Eq.~(\ref{Eq. Two-loop}). Similarly, the exponent $\eta$ is computed via the exact relation (\ref{Eq. exact eta and z}). For the special case of $\sigma=1$, the above exponents have been computed in Ref.~\cite{Sachdev04},
\begin{equation}\nonumber
  z=\frac{1}{2}+\frac{(N+2)\left(12-\pi ^2\right) }{16 (N+8)^2} \,\epsilon^2 +{\cal O}(\epsilon^3),
\end{equation}
for an $N$-component Landau-Ginzburg field theory in $d<2$ spatial dimensions, where $\epsilon=2-d$. With $d=1$ dimension ($\epsilon=1$) and $N=1$ component, we find that the above result is consistent with Eq.~(\ref{Eq. Two-loop}) where we should substitute $\sigma=1$ and $\epsilon=3\s/2-1=1/2$.

To find the behavior of our model away from the critical point, we note that the renormalization group does not change the power laws obtained at the quadratic order, specifically, logarithms such as the one in Eq.~(\ref{Eq. log b}) do not appear.

Finally, we make the comparison between our results up to the one-loop order with Ref.~\cite{Amit01}. A different renormalization scheme has been adopted by the authors of Ref.~\cite{Amit01} where they have also integrated over a frequency shell, while we have integrated over the whole range of $\omega$ at each step of RG. Therefore, their RG equations seem different from ours in Eq.~(\ref{RG for r and u}); however, we have checked that the linearized equation near the  fixed point, Eq.~(\ref{Eq. delta u delta r}), is indeed the same in both cases. While our exponent $\nu$ is different from the one in Ref.~\cite{Amit01}, we believe that the reason is a typographical error in Ref.~\cite{Amit01}. Furthermore, the exponent $\eta$ in Ref.~\cite{Amit01} associated with the spatial correlation function seems---although not explicitly stated---to be different from our definition: the definition of Ref.~\cite{Amit01} is $G(R)\sim 1/R^{d+\eta-z}$, compared to our definition $G(R)\sim 1/R^{d-1+\eta}$.

\end{widetext}

\end{document}